\documentclass[a4paper,openany, 12pt]{book}
\usepackage[text={15.5cm,23cm},centering]{geometry}

\newcommand\savemathcal[1]{%
  \expandafter\newsavebox\csname mc#1content\endcsname%
  \expandafter\savebox\csname mc#1content\endcsname{$\mathcal{#1}$}%
  \expandafter\newcommand\csname mc#1\endcsname{%
    \expandafter\usebox\expandafter{\csname mc#1content\endcsname}}%
}
\newcommand\altmathcal[1]{\csname mc#1\endcsname}
\savemathcal{Z}
\savemathcal{B}

\usepackage{mathptmx}
\usepackage{graphicx}
\usepackage{chapterbib} %remember to bibtex each chapter before final compilation
\usepackage{subfigure}
\usepackage{color}
\usepackage{amsmath}
\usepackage{amssymb}
\usepackage[colorlinks=true, linkcolor=blue]{hyperref}
\usepackage[english]{babel}  %For Vibor's institution characters

\begin{document}

\chapter{Inference from the 21cm signal}

\begin{bf}
Bradley Greig \\
\\
School of Physics, University of Melbourne, Parkville, VIC 3010, Australia \\
ARC Centre of Excellence for All-Sky Astrophysics in 3 Dimensions (ASTRO 3D) \\
\\
\noindent
Abstract
\end{bf}

In the previous chapters we have discussed in-depth the astrophysical and cosmological information that is encoded by the cosmic 21-cm signal. However, once we have a measurement, how do we extract this information from the signal? This chapter focusses on the inference of the interesting astrophysics and cosmology once we obtain a detection of the 21-cm signal. 

Essentially, inference of the astrophysics can be broken down into three parts:
\begin{itemize}
\item[1.] \textbf{Characterisation of the observed data:} The observed 21-cm signal varies spatially as well as along the line-of-sight (frequency or redshift dimension) to provide a full three dimensional movie of the intergalactic medium in the early Universe. However, we cannot perform a full pixel-by-pixel comparison between theoretical models and the observed signal. Instead, we require a variety of statistical methods to average the observational data in order to be able to better characterise and compare the behaviour of the faint signal. 
\item[2.] \textbf{An efficient method to model the 21-cm signal:} In order to interpret the observations and understand the astrophysical processes responsible, we must be able to produce physically motivated models capable of replicating the signal. Further, these must be as computationally efficient as possible in order to be able to realistically investigate the 21-cm signal.
\item[3.] \textbf{A robust probabilistic framework to extract the physics:} The observed 21-cm signal is dependent on numerous physical processes, which within our models or simulations are described by many unknown parameters. Further, these contain approximations in order to deal with the requisite dynamic range. We must be able to characterise our ignorance in a meaningful way in order to be truly able to infer the astrophysical processes of the epoch of reionisation and cosmic dawn.
\end{itemize}
\noindent
In this chapter we will focus on each separately, discussing the current state-of-the-art in inferring astrophysical and cosmological information from the 21cm signal.

\section{What do we actually measure?}

The 21-cm signal from the neutral hydrogen in the intergalactic medium is measured by its brightness temperature, $T_{\rm b}$. However, this cannot be measured directly, instead it is expressed as a brightness temperature contrast, $\delta T_{b}$, relative to the Cosmic Microwave Background (CMB) temperature, $T_{\rm CMB}$ \cite{Furlanetto:2006a}: 
\begin{equation} \label{eq:delTb}
\delta T_{b}(\mathbf{x},\nu) \equiv T_{\rm b}(\mathbf{x},\nu) - T_{\rm CMB, 0}.
\end{equation}
As such, this brightness temperature contrast can be seen either in emission or absorption, dependent on the 21-cm brightness temperature which itself is dependent on the excitation state of the neutral hydrogen (i.e. its spin temperature, $T_{\rm S}$, see Section~1.2). We can re-express Equation~\ref{eq:delTb} in terms of $T_{\rm S}$ to recover,
\begin{equation} \label{eq:delTb_Ts}
\delta T_{b}(\mathbf{x},\nu) \equiv \frac{T_{\rm S}(\mathbf{x},\nu) - T_{\rm CMB}(z)}{1+z} \left( 1 - {\rm e}^{-\tau_{\nu_{0}}(\mathbf{x},\nu) } \right),
\end{equation}
where $\tau_{\nu_{0}}$ is the optical depth of the 21-cm line (see e.g. Section~1.1). $\delta T_{b}(\mathbf{x},\nu)$ varies spatially due to its two-dimensional angular position on the sky while it varies along the line-of-sight direction owing to the 21-cm line being redshifted by cosmological expansion (i.e. adding a frequency or time dependence to the signal). Thus, measuring $\delta T_{b}(\mathbf{x},\nu)$ can reveal a full three-dimensional movie of the neutral hydrogen in the early Universe.

Unfortunately, $\delta T_{b}(\mathbf{x},\nu)$ is faint. Further, in reality it is buried under numerous astrophysical foregrounds all of which are orders of magnitude brighter (see e.g. Chapter~6). In order to deal with this faint signal coupled with the astrophysical foregrounds, typically we seek to compress the data to boost the signal-to-noise or specifically tailor methods to extract the faint signal. In Section~\ref{sec:methods}, we will discuss the numerous methods proposed in order to tease out the faint astrophysical signal from the noise.

\section{Optimal methods for characterising the 21-cm signal} \label{sec:methods}

The first step in our efforts to be able to infer information about the astrophysical processes responsible for reionisation and the cosmic dawn is to explore optimal methods to characterise the 21-cm signal. In this section, we summarise the wide variety of approaches considered in the literature, highlighting the leverage that each is able to provide with respect to the underlying astrophysical processes. Note that throughout this chapter, all investigations into detecting the 21-cm signal are generated theoretically, either analytically or numerically. Thus, we urge the reader to refer to the corresponding references in order to understand the limiting assumptions.

\subsection{Global signal} \label{sec:global}

The simplest way to deal with such a faint signal is to average it over as large a volume as possible. Since the 21-cm signal is visible across the entire sky, one can produce a complete sky-averaged (global) 21-cm brightness temperature as a function of frequency (redshift). 

\begin{figure}[]
\begin{center}
\includegraphics[trim = 0.2cm 0.6cm 0.2cm 0.2cm, scale = 0.42]{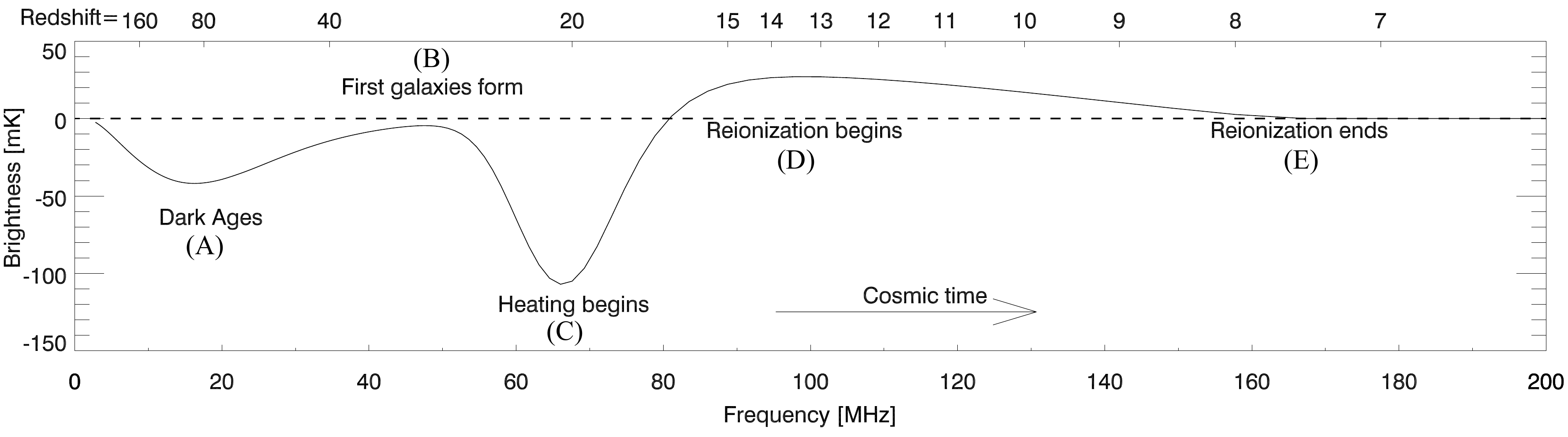}
\end{center}
\caption{A representative example of the all-sky averaged (global) 21-cm brightness temperature signal, demarcating the major cosmological transitions. Reproduced from \cite{Pritchard:2012}. \textcopyright IOP Publishing Ltd. All rights reserved.}
\label{fig:global}
\end{figure}

Although the two-dimensional spatial information from the 21-cm signal is lost, the main advantage is that it is relatively cheap to observe, requiring comparatively simple instrumentation (see e.g. Section 8.3). For example, a single radio dipole is capable of seeing essentially the entire sky at any one time, which has formed the basis for several single dipole experiments to measure the 21-cm signal. In Figure~\ref{fig:global}, we show a representative model of the global 21-cm signal, highlighting the major cosmological milestones that have been discussed in previous chapters. Thus, in each frequency bin, we measure an all-sky average of the 21-cm brightness temperature.

The global signal has been studied extensively in the literature (see e.g. \cite{Furlanetto:2006c,Pritchard:2010a,Pritchard:2012,Mirocha:2013,Fialkov:2014b,Mirocha:2014,Mirocha:2015,Mirocha:2017,Cohen:2017,Fialkov:2018,Mirocha:2018,Fialkov:2019b}. Roughly speaking, the global 21-cm signal can be broken up into five major turning points (e.g. \cite{Furlanetto:2006b,Pritchard:2010a}) corresponding to: (A) a minimum during the dark ages where collisional coupling becomes ineffective, (B) a maximum at the transition from the dark ages to the Ly$\alpha$ pumping regime (Ly$\alpha$ pumping from the first sources becomes efficient), (C) a minimum at the commencement of X-ray heating taking the signal back towards emission, (D) a maximum once the 21-cm signal becomes saturated during the EoR and finally (E) when reionisation is complete. Importantly, both the amplitude of the 21-cm signal as well as the frequency (redshift) of these transitions is strongly dependent on the underlying astrophysical processes. Thus, measuring both the amplitude and frequency of the turning points can reveal information into the underlying astrophysics.

The second turning point (end of the dark ages) can, under certain simple assumptions, be used to place limits on the spin temperature, $T_{\rm S}$. Details on $T_{\rm S}$, through equations~2.2-2.3 can provide an estimate on the overall amplitude of the angle-averaged intensity of Ly$\alpha$ photons, $J_{\alpha}$. 
The relative depth of the third turning point (heating epoch) can be used to place limits on the co-moving heating rate density, that is, the amount of heating that the IGM has undergone owing to heating sources (e.g. X-rays from HMXBs, the ISM or other more exotic scenarios. See e.g. Sections 1.3 and~2.2 for further details). Finally, if the spin temperature saturates ($T_{\rm S} >> T_{\rm CMB}$) during the epoch of reionisation then the expression for the brightness temperature (Equation~1.8) collapses into an approximate proportionality ($T_{\rm S} \propto x_{\rm H{\scriptstyle I}}(1+\delta_{\rm nl})$) with the underlying ionisation fraction, $x_{\rm H{\scriptstyle I}}$. Tracking the evolution of the ionised fraction, i.e. the reionisation history, reveals the time-span of reionisation and the number density of ionising photons produced. 

Unfortunately, the estimates for the amplitude of the ionising, Lyman-$\alpha$ and X-ray backgrounds from the global signal cannot directly reveal insights into the population of sources responsible (e.g. their typical emission spectra) as these amplitudes are convolved with the underlying galaxy number density. In compressing the entirety of the signal down into these five turning points we cannot separate out the two contributions. However, this degeneracy can be broken when further spatial information is used (e.g the 21cm power spectrum; Section~\ref{sec:PS}).

\begin{figure}[]
\begin{center}
\includegraphics[trim = 0.2cm 0.6cm 0.2cm 0.2cm, scale = 0.45]{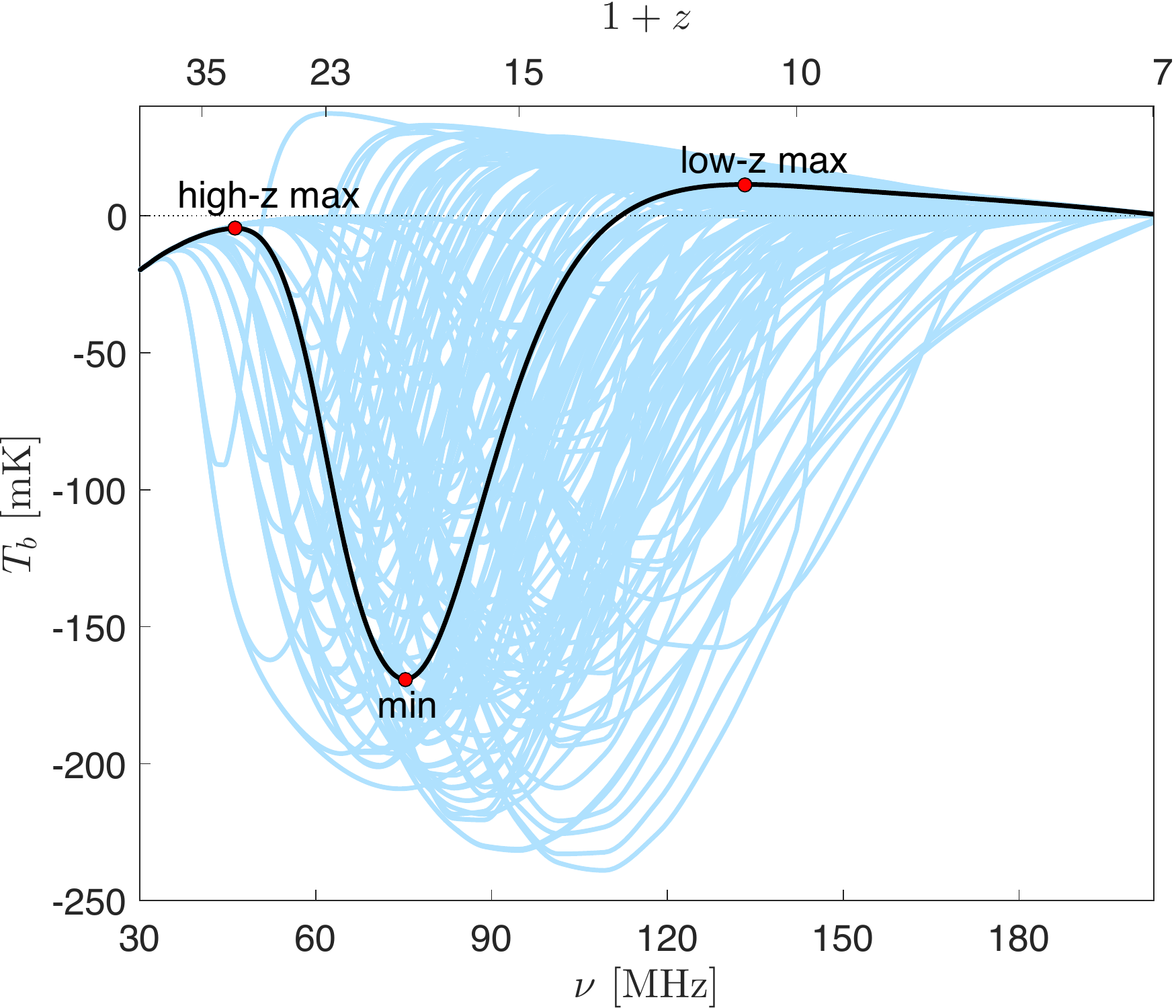}
\end{center}
\caption{The all-sky averaged (global) 21-cm brightness temperature signal obtained when varying the astrophysical parameters in $\sim~200$ theoretical models. Reproduced from \cite{Cohen:2017}. Copyright of OUP Copyright 2019.}
\label{fig:global_vary}
\end{figure}

To highlight the expected variation in the global 21-cm signal as a result of the underlying astrophysical processes, in Figure~\ref{fig:global_vary} we show $\sim~200$ theoretical models of the global 21-cm signal from \cite{Cohen:2017}. Here, the authors explore the maximal variation in the global 21-cm signal when varying the ionisation and heating properties of the astrophysical sources. Some common features in the signal are, the depth of the absorption trough deepens for lower X-ray luminosities (including some models which never appear in emission as a result of inefficient heating) or the turning points push to later times when the minimum masses of sources increases (i.e. require more massive haloes in which stars can form and produce ionising photons).

\subsection{Power spectrum} \label{sec:PS}

After the global signal, the next simplest and most straightforward approach to characterise the 21-cm signal is through the power spectrum. This is the Fourier transform of the 2-point correlation function. Basically, a measure of the excess signal (above random) on all possible spatial scales. The workhorse statistic for any signal containing structural information, the power spectrum is simply the number of modes (in Fourier space) as a function of physical scale (or size). It produces a distribution of modes characterising the amount of structural information which is contained within the signal. The power spectrum is the natural method for observing the 21-cm signal from a radio interferometer, since these measure differences in the arrival times of the cosmological signal between radio dipoles or dishes of some fixed separation. Thus, a radio interferometer is sensitive to the spatial fluctuations rather than the total amplitude.

To obtain the 21-cm power spectrum, we normalise the 21-cm brightness temperature, $\delta T_{b}(\mathbf{x})$ to be a zero-mean quantity, $\delta_{21}(\mathbf{x}) = (\delta T_{b}(\mathbf{x}) - \delta\bar{T_{b}})/\delta\bar{T_{b}}$, which amplifies the fluctuations (spatial information) in the signal. The power spectrum, $P_{21}(\mathbf{k})$ is computed by the angle-averaged sum of the Fourier transform of the 21-cm brightness temperature fluctuations via,
\begin{eqnarray}
\langle \delta_{21}(\mathbf{k}_{1})\delta^{\ast}_{21}(\mathbf{k}_{2}) \rangle = (2\pi)^{3}\delta_{D}( \mathbf{k}_{1} - \mathbf{k}_{2})P_{21}(\mathbf{k}_{1}),
\end{eqnarray}
where $\delta_{D}$ is the Dirac delta function, $\langle \rangle$ denotes the ensemble average and $^\ast$ corresponds to the complex conjugate. Typically, the 21-cm power spectrum is converted into a dimensionless quantity through $\Delta^{2}_{21}(\mathbf{k}) = (k^{3}/2\pi^{2})P_{21}(\mathbf{k})$. Typically, the Fourier modes are then averaged in spherical shells to obtain the spherically averaged power spectrum, $P_{21}(k)$, which considerably improves the overall signal-to-noise, at the cost of averaging over some spatial information. Alternatively, one can also measure the two-dimensional cylindrically averaged power spectrum, $P_{21}(k_\parallel,~k_\perp)$ decomposing it into modes perpendicular to the line-of-sight ($k_\perp$; spatially averaging the two dimensional angular modes on the sky in annuli) and along the line-of-sight ($k_\parallel$; in frequency) direction. The strength of the two dimensional 21-cm power spectrum is that most of the contamination of the signal by the astrophysical foregrounds can be contained in what is referred to as the EoR `wedge' while the remaining Fourier modes can be clean tracers of the cosmological signal (see Section~6.2.1.2).

The advantage of the power spectrum over the global signal, is that it provides a measure of the spatial fluctuations in the 21-cm signal. However, it does not encode all the available spatial information from the 21-cm signal. If these fluctuations were truly Gaussian, the power spectrum would contain all the information, and any higher order $n$-point correlation functions would contain no additional information. The structural complexity of the large and small scale processes of reionisation and the cosmic dawn results in the signal being highly non-Gaussian. As such, the power spectrum does not reveal all available information, meaning there is further constraining power from the higher order $n$-point statistics. In section~\ref{sec:bispectrum} and~\ref{sec:trispectrum}  we will return to this. Nevertheless, the power spectrum still contains a wealth of information, and observationally is considerably easier to measure.

\begin{figure}[]
\begin{center}
\includegraphics[trim = 0.2cm 0.6cm 0.2cm 0.2cm, scale = 0.4]{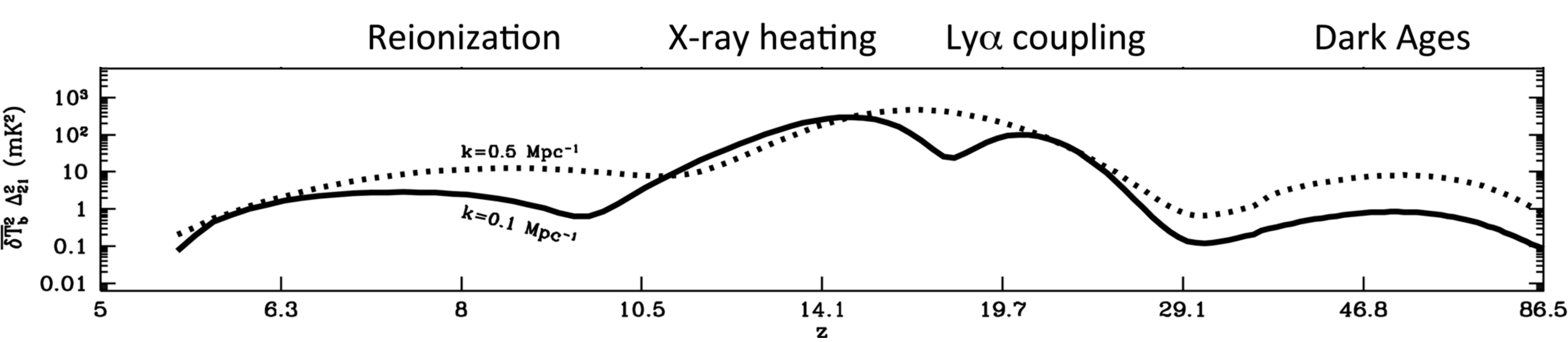}
\end{center}
\caption{The 21-cm power spectrum amplitude for two different Fourier modes, $k=0.1$~Mpc$^{-1}$ (solid) and $k=0.5$~Mpc$^{-1}$ (dashed). Peaks in the 21-cm power spectrum amplitude correspond to the different cosmic milestones. Reproduced from \cite{Mesinger:2016}. Copyright of OUP Copyright 2019.}
\label{fig:PSMilestones}
\end{figure}

The sensitivity of the 21-cm power spectrum to the underlying astrophysics can be highlighted when we decompose the 21-cm brightness temperature fluctuations through a perturbative analysis (i.e. Taylor expansion) from which we recover the following (see e.g. \cite{Barkana:2005,Santos:2005,Mao:2008}),
\begin{eqnarray}
\delta_{21} \propto C_{b}\delta_{b} + C_{x}\delta_{x} + C_{\alpha}\delta_{\alpha} + C_{T}\delta_{T} - \delta_{\partial v},
\end{eqnarray}
Simply put, fluctuations in the 21-cm brightness temperature field, $\delta_{21}$, are driven by a sum of contributions from the underlying density field, $\delta_{b}$, the ionisation fraction $\delta_{x}$, the Ly$\alpha$ coupling co-efficient, $\delta_{\alpha}$, the temperature of the neutral hydrogen $\delta_{T}$ and line-of-sight peculiar velocity gradient, $\delta_{\partial v}$. Computing the power spectrum then measures the combined signal from the power spectra of each field as well as the cross-power spectra of each. Thus, if we measure the 21-cm power spectrum across cosmic time, we will be sensitive to the epochs when each component dominates (similar to the global signal) and also the spatial scales on which the signal is strongest. This, similar to the global signal, is depicted in Figure~\ref{fig:PSMilestones}.

However, rather than using one single Fourier mode, we have a range of spatial scales over which to recover astrophysical information. This provides access to both the small-scale and large-scale physical processes. For example, during the EoR, the 21-cm power spectrum is dominated by the contribution from the ionisation field, which contains particular structural information on the reionisation process due to the characteristic size of the H$_{\rm \scriptsize II}$ regions as well as their clustering (e.g. \cite{Lidz:2008}). One can equally obtain the spectrum of the sources responsible for heating the IGM from the structural information, owing to the strong dependence of the mean free path with the energy of the X-ray sources. 

\begin{figure}[]
\begin{center}
\includegraphics[trim = 0.2cm 1cm 0.2cm 0.2cm, scale = 0.75]{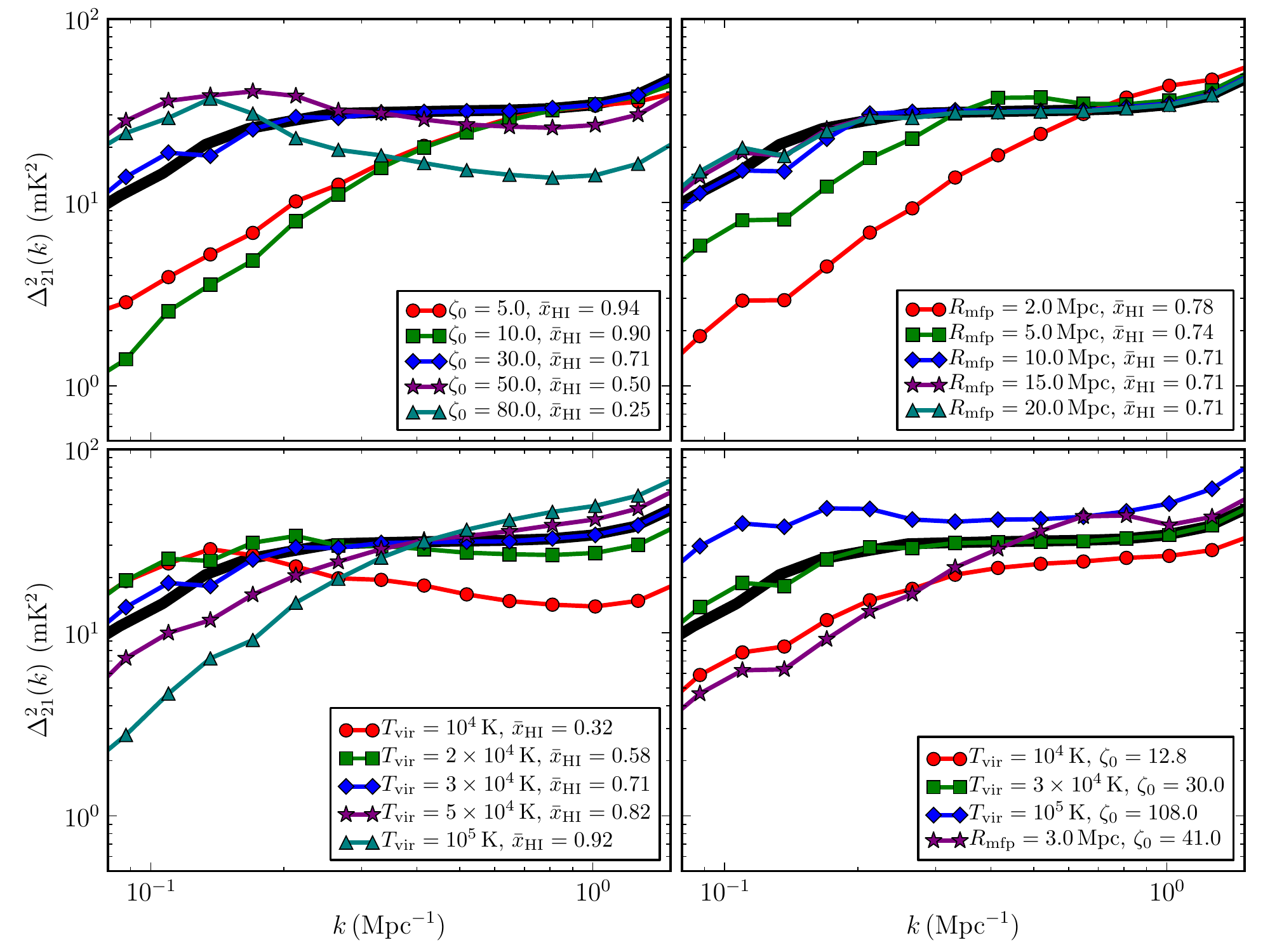}
\end{center}
\caption{The three dimensional spherically averaged 21-cm power spectrum at $z=9.0$ when varying astrophysical parameters controlling different astrophysical processes, assuming $T_{\rm S} \gg T_{CMB}$. Top left: the number of ionising photons produced per baryon (ionising efficiency, $\zeta$), top right: maximum ionising photon horizon (proxy for maximum allowable bubble size, $R_{\rm mfp}$) and bottom left: minimum mass of halo hosting star-forming galaxy (represented here as $T_{\rm vir}$). Bottom right: several models at the same ionisation fraction. Reproduced from \cite{Greig:2015}. Copyright of OUP Copyright 2019.}
\label{fig:PSvariation}
\end{figure}

In Figure~\ref{fig:PSvariation} we show the variation in the three dimensional spherically averaged 21-cm power spectrum at a single redshift ($z=9$) when varying three different astrophysical parameters under the assumption of $T_{\rm S} \gg T_{CMB}$~(see e.g. \cite{Greig:2015}). Inset tables correspond to the parameter being varied and the resultant IGM neutral fraction (stage of reionisation). In the top left panel, we vary the ionising efficiency, $\zeta$, a proxy for the number of ionising photons produces by the sources. The shape of the 21-cm power spectrum differs considerably with ionising efficiency. In the early stages, the 21-cm PS matches the density (matter) power spectrum, while in the latter stages it follows the ionisation field. 

Similar behaviour is observed for varying $T_{\rm vir}$, a proxy for the minimum mass of halos hosting star-forming galaxies. Increasing this threshold, results in fewer sources to contribute to reionisation. In the top right panel, the maximum photon horizon, $R_{\rm mfp}$, is varied. Essentially, in this specific work it acts as a maximum allowable bubble size. Note that in this case, the change in $R_{\rm mfp}$ does not alter the neutral fraction strongly, thus the changes in the 21-cm power spectrum are purely as a result in changes to the size of the ionised regions. Finally, in the bottom right we highlight astrophysical models with the same IGM neutral fraction (i.e. the same stage of reionisation). Despite being at the same point in reionisation, the amplitude and shape of the 21-cm power spectrum differs considerably, highlighting the sensitivity of the 21-cm power spectrum to the underlying astrophysical parameters. 

While this example is only for the epoch of reionisation, the same strong sensitivity of the 21-cm power spectrum to the underlying astrophysics is true for both the heating or Ly$\alpha$ coupling epochs. This highlights the strength and utility of the 21-cm power spectrum for recovering the astrophysical information. As such numerous authors have explored the impact of various astrophysical processes on the 21-cm power spectrum (see e.g. \cite{Bowman:2006,Furlanetto:2006a,Iliev:2006,McQuinn:2006,McQuinn:2007,Pritchard:2007,Lidz:2008,Santos:2008,Baek:2010,Harker:2010,Mesinger:2013,Fialkov:2014b,Pober:2014,Greig:2015,Geil:2016,Greig:2017b,Hassan:2017,Cohen:2018,Greig:2018,Park:2019,Seiler:2019}).

\subsection{Bispectrum} \label{sec:bispectrum}

The logical extension beyond the power spectrum, the bispectrum, $B$, is simply the Fourier transform of the 3-point correlation function,
\begin{eqnarray}
\langle \delta_{21}(\mathbf{k}_{1})\delta_{21}(\mathbf{k}_{2})\delta_{21}(\mathbf{k}_{3}) \rangle = (2\pi)^{3}\delta_{D}( \mathbf{k}_{1} - \mathbf{k}_{2} - \mathbf{k}_{3})B(\mathbf{k}_{1},\mathbf{k}_{2},\mathbf{k}_{3}),
\end{eqnarray}
where the $\delta_{D}$ enforces that the Fourier modes must form closed triangles. It measures the excess probability of the underlying quantity as a function of three spatial positions in real space. The bispectrum provides a scale-dependent measure of the non-Gaussianity of the 21-cm signal, and as such contains additional astrophysical information beyond that held in the power spectrum. However, it suffers from lower signal-to-noise as there are less modes to average over to boost the signal.

Whereas the power spectrum is relatively trivial to interpret as it is a measure of the power over a single length scale, $k$, the bispectrum is the measure of power over all possible triangle configurations that satisfy the closure condition from $\delta_{D}$. Thus in order to simplify the interpretation of the bispectrum, it is common to consider several simplified triangle configurations. These are typically: (i) the equilateral triangle ($k_{1}  = k_{2} = k_{3}$), (ii) the isosceles triangle ($k_{1} > k_{2} = k_{3}$), (iii) folded triangle ($k_{1} = 2k_{2} = 2k_{3}$), (iv) elongated triangle ($k_{1} = k_{2} + k_{3}$) and (v) the squeezed triangle ($k_{1} \simeq k_{2} \gg k_{3}$). Each, corresponds to different physical properties of the real-space field.

While a detailed discussion of the 21-cm bispectrum is beyond the scope of this chapter, it is fruitful to provide a brief explanation and example of the various configurations (see for example \cite{Lewis:2011} and \cite{Watkinson:2019} for more detailed discussions). The equilateral configuration is essentially an extension of the power spectrum, in the sense that it is expressed as a single amplitude scale, $k$. Generally speaking, it produces the largest amplitude signal and as such is the most commonly studied configuration. It is sensitive to the spherical symmetry of the 21-cm signal such as the scale of the ionised H$_{\rm \scriptstyle II}$ regions during reionisation or the hot/cold spots due to IGM heating. Typically its amplitude grows during the EoR as the signal becomes more non-Gaussian due to the topology of the ionisation field. Shifting towards isosceles or folded triangles, these become more sensitive to planar or filamentary structures in the underlying 21-cm signal. Thus as the topology of either the ionised or X-ray heated regions deviate away from spherical symmetry (i.e. either multiple contributing sources or overlap of ionised regions) the signal should increase with increasing angle. The squeezed limit correlates the small-scale signal from two modes with a large-scale mode, for example capturing the impact of the large-scale environment (i.e. from X-ray heating) on the small-scale power spectrum (i.e. source clustering).

In addition to the structural information in the bispectrum amplitude, the relative sign of the bispectrum under certain triangle configurations and on certain spatial scales can equally reveal insights into the underlying processes. As discussed in \cite{Majumdar:2018,Hutter:2019}, the sign of the bispectrum during reionisation can help distinguish between whether the non-Gaussianity is driven by the topology of the ionised regions (where the bispectrum is negative owing to the below average contribution from the ionised regions) compared to being driven by the matter and cross-bispectra (where it is positive). 

\begin{figure}[]
\begin{center}
\includegraphics[trim = 0.2cm 1cm 0.2cm 0.2cm, scale = 0.52]{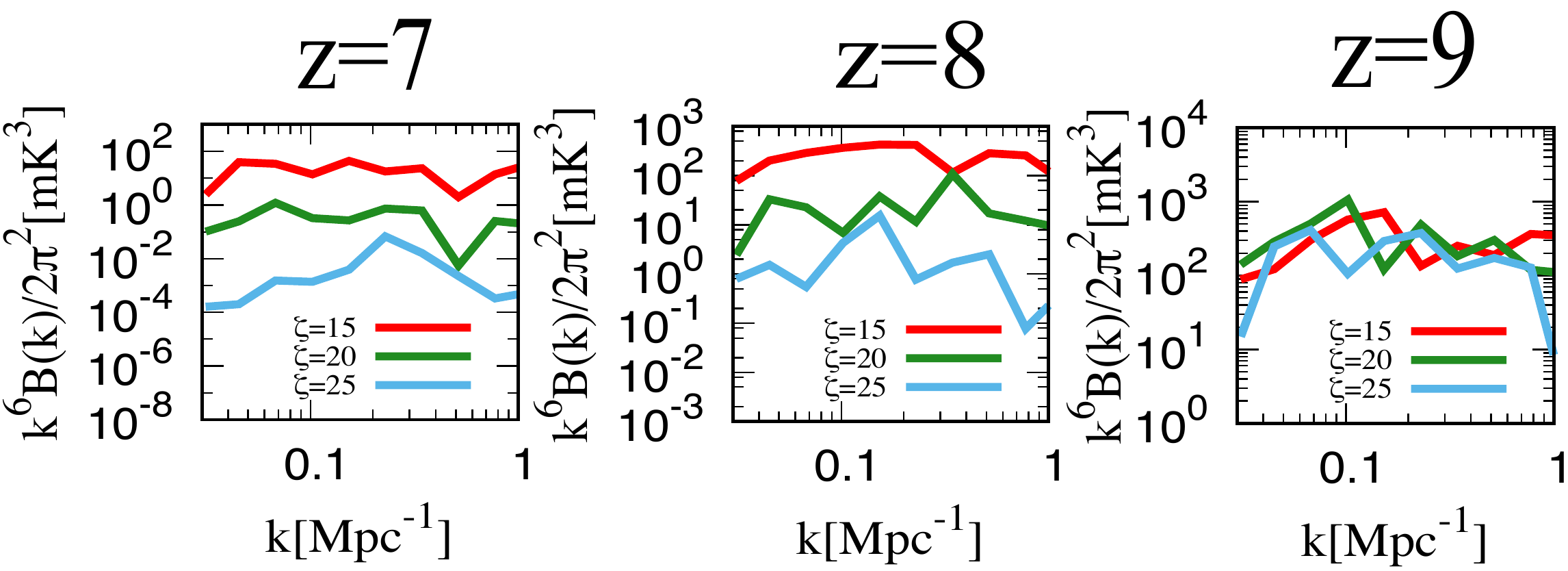}
\end{center}
\caption{Variation in the amplitude of the equilateral bispectrum at $z=7$, 8 and 9 for different ionising efficiencies, $\zeta$. Reproduced from \cite{Shimabukuro:2017}. Copyright of OUP Copyright 2019.}
\label{fig:EqBS}
\end{figure}

In Figure~\ref{fig:EqBS}, we compare the equilateral bispectrum at $z=7$, 8 and 9 from \cite{Shimabukuro:2017} for differing ionising efficiency, $\zeta$. For decreasing $\zeta$, the amplitude of the bispectrum increases due to its amplitude being dependent on the ionisation fraction. Thus, different reionisation models are easily distinguishable by the 21-cm bispectrum. 

In recent times, the 21-cm bispectrum has gained considerable traction in interpreting the astrophysics of reionisation and the cosmic dawn (see e.g. \cite{Bharadwaj:2005,Pillepich:2007,Yoshiura:2015,Shimabukuro:2016,Shimabukuro:2017,Watkinson:2017,Majumdar:2018,Hutter:2019,Trott:2019,Watkinson:2019}). Alternatively, rather than exploring the information from the amplitude of the Bispectrum, \cite{Gorce:2019} introduced a three point correlation function based solely on the phases of the Fourier modes (e.g.~\cite{Obreschkow:2013}), termed the triangle correlation function. In focussing solely on the phases, it is sensitive to the characteristic size of the ionised regions and thus exploring the topology of reionisation, which places it in a similar vein as methods to other topological based approaches (Section~\ref{sec:topology}) or the size distribution of ionised regions (Section~\ref{sec:BSDs}). However, not all experiments are designed to measure this phase information. In fact, several experiments are specifically designed to throw away this phase information for increased sensitivity to specific spatial scales. These are referred to redundant configurations and are discussed in Chapter 7.

\subsection{Trispectrum} \label{sec:trispectrum}

Following the Bispectrum, the Trispectrum is the Fourier transform of the four-point correlation function. Already at the level of the Bispectrum, the relative signal-to-noise of the signal is becoming weak, thus, in the foreseeable future it is unlikely a measurement of the Trispectrum during the EoR or earlier will be achievable. Nevertheless, \cite{Cooray:2008} explored the Trispectrum of the 21-cm fluctuations, focussing on fundamental cosmology rather than the astrophysics of the reionisation process. These authors find that the anisotropies from the 21-cm signal are sensitive to primordial non-Gaussianities, an important quantity in constraining inflationary models.

\subsection{One-point statistics}

Rather that measuring the Fourier transform (e.g. power spectrum) of the 21-cm brightness temperature signal, $\delta_{21}(\mathbf{x})$, we can instead measure the one-point statistics (or moments) of the probability distribution function (PDF). In fact, we have already discussed the lowest order one-point statistic, that is, the mean of $\delta T_{b}(\mathbf{x})$ given by the global signal (see \ref{sec:global}).  These one-point statistics of the PDF essentially measure the deviations away from a fully Gaussian PDF, thus they are by definition sensitive to the non-Gaussian nature of the 21-cm signal. Generally speaking, the one-point statistics of $\delta T_{b}(\mathbf{x})$ are given by,
\begin{eqnarray}
m_{n} = \frac{1}{N}\sum^{N}_{i=0}(\delta T_{b}(\mathbf{x_{i}}) - \bar{\delta T_{b}})^{n},
\end{eqnarray}
where $m_{n}$ is the $n$-th order moment and $N$ is the number of pixels over which the signal is measured. For the 21-cm signal, these moments would be generated from the observed two-dimensional tomographic maps of the 21-cm signal.

The next lowest order statistic of the PDF following the mean is the variance, $\sigma^{2}$. The variance is equivalent to the average of the power spectrum over all Fourier modes, $k$,
\begin{eqnarray}
\sigma^{2} = (\bar{\delta T_{b}})^{2} \int \frac{d^{3}k}{(2\pi)^{3}} P(\mathbf{k}).
\end{eqnarray}
As it is the average over all spatial information, the variance itself is less sensitive to the underlying astrophysics than the power spectrum. However, the strength of one-point statistics shines through when using the higher order moments in combination with the variance (or power spectrum). The next two higher order moments are referred to as the skewness and the kurtosis. Equivalent to the variance's relation to the power spectrum, the skewness and kurtosis are the average over all Fourier modes of the bispectrum and trispectrum respectively (the three and four-point correlation functions). As such, whereas the power spectrum only measures the 2-point correlations, the skewness and kurtosis reveals insights from the non-Gaussian properties of the 21-cm signal.

\begin{figure}[]
\begin{center}
\includegraphics[trim = 0.2cm 1cm 0.2cm 0.2cm, scale = 0.42]{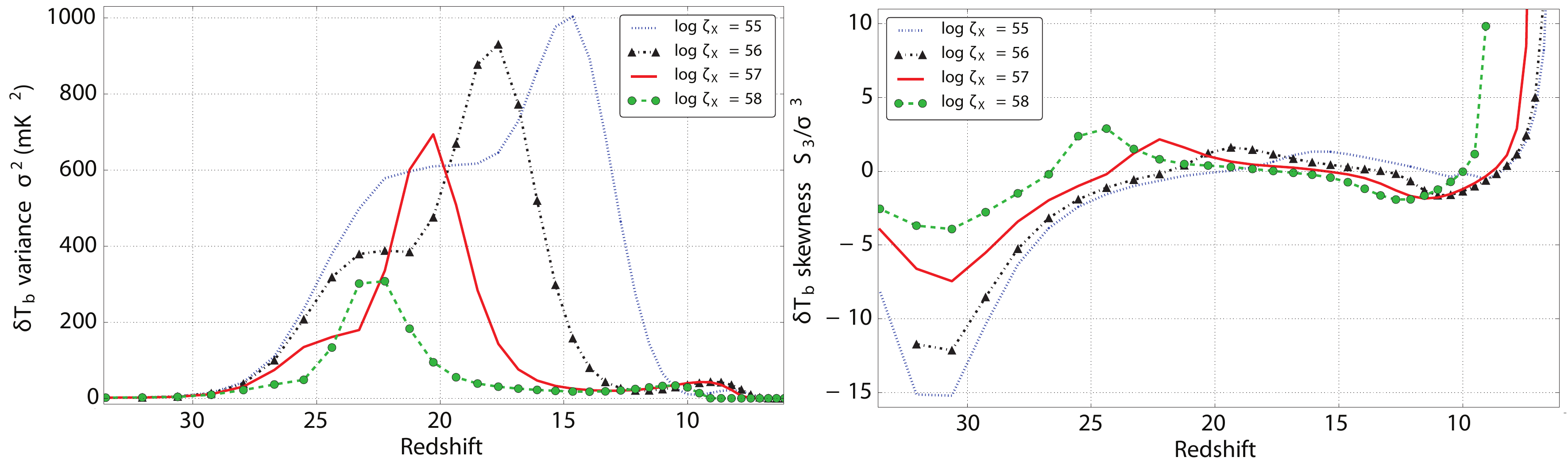}
\end{center}
\caption{The variance (left) and normalised skewness (right) of the 21-cm brightness temperature when varying the efficiency of X-ray heating in the IGM. Reproduced from \cite{Watkinson:2015}. Copyright of OUP Copyright 2019.}
\label{fig:skewness}
\end{figure}

The amplitude of the variance is sensitive to differences in the 21-cm brightness temperature. For example, during the EoR, as the number of ionised regions increases (i.e. the contrast between the 21-cm signal from the neutral regions compared to zero signal from the ionised regions) the variance increases. It subsequently turns over as most of the volume is ionised. The skewness is a measure of the asymmetry of the underlying PDF. A negative skewness corresponds to a longer tail towards a lower amplitude signal and a positive skewness corresponds to a longer tail towards higher amplitude signals. The kurtosis is essentially a measure of the outliers of the distribution, with increasing positive (negative) kurtosis corresponding to larger positive (negative) amplitude outliers.

Figure~\ref{fig:skewness} shows an example of both the variance (left) and normalised skewness (right panel) of the 21-cm brightness temperature under different levels of X-ray heating. For increasing X-ray efficiencies (i.e. increase heating) the peak of the variance decreases in amplitude while shifting to earlier times. Increasing the efficiency allows the X-ray heating to occur earlier, reducing the contrast between the $T_{\rm CMB}/T_{\rm S}$ resulting in a lower amplitude peak in the variance. This same behaviour equally results in larger skewness for decreasing X-ray efficiency, owing to a more asymmetric PDF of 21-cm brightness temperatures due to the increasing contrast in $T_{\rm CMB}/T_{\rm S}$. Clearly from Figure~\ref{fig:skewness} it can be seen that these one-point statistics are capable of distinguishing between different astrophysical models. As such, these one-point statistics have been explored in numerous works (e.g. \cite{Wyithe:2007b,Harker:2009,Patil:2014,Watkinson:2014,Watkinson:2015,Kittiwisit:2016,Kubota:2016,Watkinson:2015b,Shimabukuro:2015,Ross:2017}).

Alternatively, the direct 21-cm PDF or the difference PDF have also been studied (e.g. \cite{Barkana:2008,Gluscevic:2010,Ichikawa:2010,Pan:2012}). The difference PDF is the difference between the brightness temperature separated by some spatial scale, $r$. The advantages of the difference PDF is that it can bypass the fact that interferometric observations cannot easily determine the zero flux threshold of the 21-cm signal and that it includes more data by being dependent on spatial scales (similar to two-point correlation functions or the power spectrum). The difference PDF can be more sensitive to the ionising sources and sizes of the ionised regions as it is a direct measure of the distribution of separated pixel pairs that are either both ionised, ionised and neutral or both neutral.

\subsection{Wavelets}

Thus far we have only considered either real-space quantities such as the one-point statistics or the Fourier transform of the $n$-point correlation functions (i.e. the power spectrum and bispectrum). The Fourier transform measures the amplitude of the fluctuations of a given spatial scale, and in order to increase the signal-to-noise we must average the signal over all line-of-sight modes within some observed bandwidth. As a result, we average over modes containing different redshift evolutions and thus increase the bias of the signal. This can be minimised somewhat, for the case of the power spectrum, by averaging the signal over relatively narrow observing bandwidths. However, it still results in some loss in fidelity of the signal.

Instead, in \cite{Trott:2016} the authors explore the potential usage of wavelets, which provide multiple alternatives to the Fourier basis set. Specifically, they explored the application of the Morlet Transform. This provides a family of curves which provide the ability to localise the 21-cm signal both spatially and in frequency. The equivalent to the power spectrum, the Morlet power spectrum is capable of providing an unbiased estimator which maximises the three dimensional nature of the 21-cm signal. Preliminary analysis shows that the Morlet power spectrum performs more optimally than the Fourier power spectrum. A physical interpretation of the Morlet power spectrum in the context of the evolution of the 21-cm signal has yet to be explored.

\subsection{Topological measurements of the 21-cm signal} \label{sec:topology}

Up until this point, we have only discussed methods of characterising the 21-cm signal using just the amplitude of the spatial (e.g. Fourier) information. This is primarily driven by the difficulty in measuring the 21-cm signal and the low signal-to-noise of the first generation experiments. However, the most advanced radio interferometers (such as the Square Kilometre Array, SKA; see Section~9.2.1) should be able to provide two dimensional images of the 21-cm signal. That is, they should provide significant signal-to-noise to enable both the amplitude and phase information to be used. 

Direct images of the 21-cm signal contain the complicated morphology of the hot (above average or over dense signal) and cold (below average or under dense signal) of the 21-cm brightness temperature throughout the history of reionisation and the cosmic dawn. The relative sizes, shapes and clustering of these hot/cold patches can reveal numerous insights into the underlying astrophysical processes, such as the number density of sources, their contribution to the heating/ionisation of the IGM and the shape of the emitted spectrum of radiation. The study of these geometric shapes in mathematics is referred to as topology.

Topological studies of reionisation and the cosmic dawn are complimentary to the methods described previously. For example, reionisation proceeds through three main stages (e.g. \cite{Gnedin:2000,Furlanetto:2016}): pre-overlap, over-lap and post-overlap. In pre-overlap, the first ionised H {\scriptsize II} regions (or bubbles) grow completely in isolation roughly until $x_{\rm HI} \geq 0.1$. Over-lap ($ 0.9 \geq x_{\rm HI} \geq 0.1$) describes the merging of these ionised bubbles into essentially a single large connected ionised region. Finally, post-overlap $x_{\rm HI} \geq 0.9$ corresponds to the breaking down of the last remaining patches of neutral IGM into smaller and smaller islands. Topological studies are capable of breaking down these transitions by describing the ratios of ionised and neutral regions, how the ionised (or neutral) regions are connected together and how they are embedded in the larger structures as they form. This provides unique insights into the reionisation epoch not available from statistical methods.

Unfortunately we cannot perform a full pixel by pixel analysis of a measured 21-cm image, therefore we must still compress our images into some form of statistical measurement. There are numerous methods to attempt to characterise the topology of the 21-cm signal. Below, we summarise several of the main approaches taken in the literature. Fundamental to topological studies is the definition of how to identify regions of interest. Typically, a threshold value is required, with the quantity above/below this threshold being used to distinguish the two regions.

\subsubsection{Genus or the Euler characteristic}

The genus, $g$, is a topological property that defines the number of cuts one can make to an object (i.e. H {\scriptsize II} region) without dividing it into independent disconnected sub-regions. It can simply be expressed as,
\begin{eqnarray}
g = N_{\rm >th} - N_{\rm < th}
\end{eqnarray}
where $N_{\rm >th}$ and $N_{\rm < th}$ are the number of connected (or fully enclosed) regions above and below the threshold value for identification. By gradually increasing the threshold value from some initial starting value, a genus curve is constructed, which is a measure of the connectedness of the quantity as a function of different threshold values (e.g. $x_{H {\rm \scriptsize I}}$, $\delta T_{b}$). Typically, these threshold values are expressed in units of the standard deviation from the mean.

The genus has been explored, both in two and three dimensions, either in the context of the ionised (or neutral) field (\cite{Gleser:2006,Lee:2008,Friedrich:2011}) or the 21-cm brightness temperature field (\cite{Hong:2014,Wang:2015}). However, it has yet to be explored in the context of the heating epoch (i.e. $T_{\rm S} \gg T_{\rm CMB}$ is typically assumed). For a purely Gaussian field, the genus curve is symmetric around zero. Thus deviations from symmetry highlight the non-Gaussianity of the 21-cm signal. 

Differences in the evolution in the amplitude of the genus as a function of threshold density can distinguish different source biases and ionising efficiencies. For example, reionisation driven by larger, more biased sources exhibits a different topology than one driven by numerous fainter sources. This appears as changes in the amplitude of the genus as a function of threshold. When the ionised regions are isolated, the genus amplitude is higher than when they begin to overlap (as the total number of isolated ionised regions decreases).

\subsubsection{Minkowski functionals}

A more generalised description of the geometry or topology of the 21-cm signal can be obtained from what are referred to as Minkowski functionals. These are well known concepts from the branch of mathematics known as integral geometry. In $n$-dimensions, there exists $n+1$ independent Minkowski functionals which means that in three dimensional space we have four functionals to describe the topology. Used heavily in cosmology, in particular geometrical features of the galaxy distribution (e.g. \cite{Gott:1986,Schmalzing:1997}) and non-Gaussianity of the CMB (e.g. \cite{Komatsu:2009}), recently they have gained favour for describing the topology of reionisation \cite{Gleser:2006,Friedrich:2011,Yoshiura:2017,Chen:2018}.

For a zero mean scaler function, $u(x)$, (e.g. $\delta T_{\rm b}$) within a volume, $V$, and standard deviation, $u$, we can define an excursion set, $F_{\nu}$, which contains all points that satisfy the threshold, $u(x) \geq \nu\sigma$, where $\nu = u_{\rm th}/\sigma$ and $u_{\rm th}$ is the threshold value. Mathematically, this gives rise to the following Minkowski functionals,
\begin{eqnarray}
V_{0}(\nu) = \frac{1}{V}\int_{V} {\rm d^3}x\,\Theta\left[u(x) - \nu \sigma\right] \\
V_{1}(\nu) = \frac{1}{6V}\int_{\partial F_{\nu}} {\rm d}s \\
V_{2}(\nu) = \frac{1}{6\pi V}\int_{\partial F_{\nu}} {\rm d}s\,\left[\kappa_{1}(x) + \kappa_{2}(x)\right] \\
V_{3}(\nu) = \frac{1}{4\pi V}\int_{\partial F_{\nu}} {\rm d}s\, \kappa_{1}(x)\kappa_{2}(x).
\end{eqnarray}
Here, $\Theta$ is the Heaviside step-function, $\partial F_{\nu}$ is the surface of the excursion set, ${\rm d}s$ is the surface element and $\kappa_{1}(x)$ and  $\kappa_{2}(x)$ are the principle curvatures (inverse of the principle radii) at $x$. The zeroth Minkowski functional, $V_{0}$, corresponds simply to the total volume of the excursion set (i.e. volume above the threshold value), $V_{1}$ and $V_{2}$ correspond to the total surface and mean curvature of the excursion set while $V_{3}$ is the integrated Gaussian curvature over the surface or the Euler characteristic (also $\chi$). The Euler characteristic is related to the genus, $g$, via $V_{3} = 2(1-g)$ thus it effectively describes the shape of the excursion set. Thus, the full set of Minkowski functionals contain additional information beyond that of just the genus.

\begin{figure}[]
\begin{center}
\includegraphics[trim = 0.2cm 1cm 0.2cm 0.2cm, scale = 0.42]{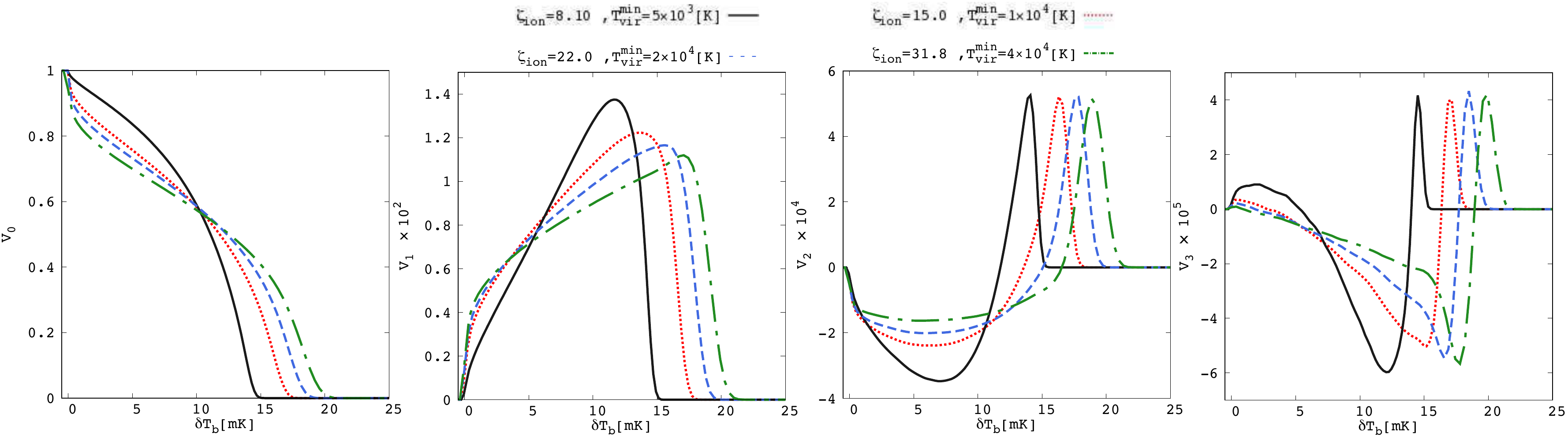}
\end{center}
\caption{The impact of varying the astrophysical parameterisation for a fixed neutral fraction ($x_{H_{\rm \scriptstyle I}}\approx 0.5$) and redshift ($z=8.6$). The coloured curves highlight the impact of varying either the ionising efficiency $\zeta$ or the minimum halo mass for star-forming galaxies, $T_{\rm vir}$ on the four Minkowski functionals. Reproduced from \cite{Yoshiura:2017}. Copyright of OUP Copyright 2019.}
\label{fig:MFs}
\end{figure}

In Figure~\ref{fig:MFs} we show the four Minkowski functionals for the 21-cm brightness temperature when varying the underlying astrophysical processes from \cite{Yoshiura:2017} at a fixed neutral fraction ($x_{H_{\rm \scriptstyle I}}\approx 0.5$) and redshift ($z=8.6$). Here, these authors consider variations in either the ionising efficiency, $\zeta$, or the minimum halo mass hosting star-forming galaxies, $T_{\rm vir}$. Clearly, different reionisation histories are distinguishable by the Minkowski functionals.

Generally speaking the following behaviour is expected of the Minkowski functionals throughout reionisation and the cosmic dawn. $V_{0}$ describes the volume contained above/below the threshold value. For example, if $V_{0} \sim 0.5$ at $\delta T_{\rm b} = 0$ this implies the number of patches above/below the average 21-cm signal are roughly equal. The $V_{0}$ curve will move from left to right (to increasing $\delta T_{\rm b}$) as heating of the IGM occurs. $V_{1}$ (reflected in $V_{2}$) exhibits a similar shift to higher $\delta T_{\rm b}$, however it is initially strongly peaked with a high density tail containing the heated regions. This peak smooths out over a broader range of $\delta T_{\rm b}$ as IGM heating continues. During reionisation, $V_{1}$, $V_{2}$ and $V_{3}$ will shift toward $\delta T_{\rm b}=0$ as the higher amplitude $\delta T_{\rm b}$ regions ionise first.

\subsubsection{Shape-finders}

An extension to Minkowski functionals, shape-finders (\cite{Sahni:1998}) are a way to characterise the shapes of compact surfaces. Applied to reionisation (\cite{Bag:2018,Bag:2019}), these shape-finders can provide a means to characterise how the ionised regions grow. For example, they are useful in being able to distinguish between whether the topology is planar or filamentary. Shape-finders are derived directly from the Minkowski functionals via:
\begin{eqnarray}
{\rm Thickness}: T = \frac{3V_{0}}{V_{1}} \\
{\rm Breadth}: B = \frac{V_{1}}{V_{2}} \\
{\rm Length}: L =  \frac{V_{3}}{4\pi}.
\end{eqnarray}
These shape-finders are interpreted as providing the three principle axes of a physical object. The morphology of the ionised region can then be defined by either the planarity or its filamentarity:
\begin{eqnarray}
{\rm Planarity}: P = \frac{B - T}{B + T} \\
{\rm Filamentarity}: F = \frac{L - B}{L+B},
\end{eqnarray}
where $P\gg F$ corresponds to planar objects (i.e. sheets) while the opposite $F\gg P$ corresponds to a filament.

During the reionisation epoch, percolation theory shows that a single infinitely large, multiply connected ionised region will rapidly form (e.g. \cite{Furlanetto:2016}). When describing the largest singly connected ionised region, \cite{Bag:2018,Bag:2019} find that both $T$ and $B$ evolve slowly whereas $L$ increases rapidly. Thus, this large ionised region grows only along its `length' implying a highly filamentary structure.

\subsubsection{Persistent homology theory}

Homology characterises the topology of the ionisation bubble network into its fundamental components: ionised regions, tunnels (enclosed neutral filaments) and cavities (patches of neutral hydrogen). The persistence then quantifies the significance of the feature, for example its lifetime, by computing a birth and death date for an object.  Thus far, it has only been applied to the ionisation field (\cite{Elbers:2019}). These ionised regions ($\beta_{0}$), tunnels ($\beta_{1}$) and cavities ($\beta_{2}$) can be described by the so-called Betti numbers, $\beta_{n}$, which contain the total number of each type of structure. These can be related to the earlier Euler characteristic via, $\chi = \beta_{0} - \beta_{1} + \beta_{2}$. By breaking the Euler characteristic into the constituent components and tracking their individual growth reveals additional information on the topology, thus it is a more generalised method than either the genus of the Minkowski functionals.

\subsubsection{Fractal dimensions}

An alternative to classifying the ionised (neutral) regions embedded in the 21-cm signal is through a fractal dimensions analysis. Applied to reionisation (\cite{Bandyopadhyay:2017}), this provides a direct means to quantify the deviation away from a homogenous distribution, as well as the degree of clustering and lacunarity (a measure of the size of the ionised regions). The fractal dimension, $D_{q}$, also known as the Minkowski-Bouligand dimension, is a measure of how complicated the topology of the field in question is. A homogeneous distribution in three dimensions has a $D_{q}=3$. \cite{Bandyopadhyay:2017} show that the topology of reionisation exhibits a significant multi-fractal behaviour. These authors find that the fractal dimension is relatively insensitive to the minimum halo mass of the star-forming galaxies, however it was sensitive to the mass averaged ionisation fraction. Thus, the correlation dimension can be useful for constraining the global neutral fraction. Additionally, it is a strong discriminant of models of outside-in and inside-out reionisation.

\subsubsection{Contour Minkowski tensor}

In \cite{Kapahtia:2018,Kapahtia:2019}, these authors introduced the rank-2 contour Minkowski tensor (e.g.~\cite{mcmullen1997,alesker1999,beisbart2002,hug2008,schroder2010,Schroder-Turk:2013}) in two-dimensions which can probe both the length and time scales of the ionised regions during reionisation. The Minkowski tensors are a generalisation of the scalar Minkowski functionals. The contour Minkowski tensor provides information on both the alignment of structures in two dimensions and their anisotropy. Since the ionised regions are not perfectly spherical, their shape anisotropy can be explored by the ratio of the two eigenvalues of the contour Minkowski tensor while the amplitude of the eigenvalues describes their size.

In this analysis, the number of connected regions and holes (e.g. the Betti numbers) given a specific threshold value are tracked. In addition, a characteristic radius of the structures and their shape anisotropy can be determined. For a description of the evolution of $\delta T_{\rm b}$, we refer the reader to \cite{Kapahtia:2019}, ignoring it here owing to its complexity due to the definition of the connected regions and holes as a function of the threshold value as the 21-cm signal transitions transition from above/below average signal regions in the heating epoch to neutral/ionised regions during reionisation. However, we emphasise that these authors explored varying the minimum mass hosting star-forming haloes and clearly show that different astrophysical parameters can be distinguishable.

\subsection{Bubble size distributions} \label{sec:BSDs}

Throughout reionisation and the cosmic dawn, the morphology of the 21-cm signal is driven by processes that embed a morphological signature on the 21-cm signal. For example, the ionised H$_{\rm \scriptsize II}$ regions or the hot (above average signal) or cold (below average signal) spots in the 21-cm brightness temperature during the heating epoch. Quite simply, if we could measure the distribution of these `bubbles' and how they evolve over cosmic time we would have a strong discriminant of the populations of sources responsible for the heating and ionisation of the IGM and also the spectrum of their emitted radiation. Effectively, this would behave as a statistical distribution function (number of bubbles given a physical scale) analogous to a halo mass function. However, the bubbles do not remain isolated, very quickly overlapping into increasingly large and topologically complex structures. Thus, there is no unique way to characterise these bubbles. Nevertheless several methods have been explored in order to be able to construct a probabilistic distribution of the bubble sizes.

The simplest is a friends-of-friends approach (e.g. \cite{Iliev:2006,Friedrich:2011}), which simply connects all cells above (below) a threshold value. However, very rapidly a single large ionised structure exists which fills most of the volume with only a small fraction of isolated regions remaining. The relative volume of this large ionised region and the distribution of the smaller regions can still differentiate reionisation morphologies, however it contains less statistical weight. Alternatively, in \cite{Zahn:2007} a sphere is placed on every pixel, averaging the signal across increasingly larger spheres until a radius is found where the average signal is above the threshold value. While this generates a more statistical meaningful distribution of bubbles, these sizes tend to overestimate the size of the topological feature of interest due to the assumed spherical symmetry.

Recently, more statistically robust methods have been introduced to measure the bubble size distributions. First of these is the mean free path method, which uses a Monte Carlo approach by considering a large number of random positions and determining the distance to the edge of the bubble from different random directions (e.g. \cite{Mesinger:2007}). This results in an unbiased estimator of the bubble size distribution (e.g. \cite{Lin:2016}). 

The Watershed method (e.g. \cite{Lin:2016}) is a more sophisticated approach and has been readily used in the search for cosmological voids. It is a well known two-dimensional image segmentation algorithm creating contours of constant value (i.e. $\delta T_{b}$) which are treated as levels of a tomographic map. These are then `flooded' to obtain unique locations for the minima (e.g. ionised regions). Remaining in the image processing regime, \cite{Giri:2018} introduced the superpixels method. This uses a region based method to identify regions of complex shapes (i.e. ionised regions) segmenting these regions into smaller segments called superpixels. The bubble size distribution is then obtained by averaging the value of the 21-cm brightness temperature within each superpixel before constructing the PDF. Finally, granulometry \cite{Kakiichi:2017} has been investigated, which effectively performs a series of sieving operations to construct a distribution of the sizes of objects which pass through sieves of various sizes and shapes. 

\begin{figure}[]
\begin{center}
\includegraphics[trim = 0.2cm 1cm 0.2cm 0.2cm, scale = 0.65]{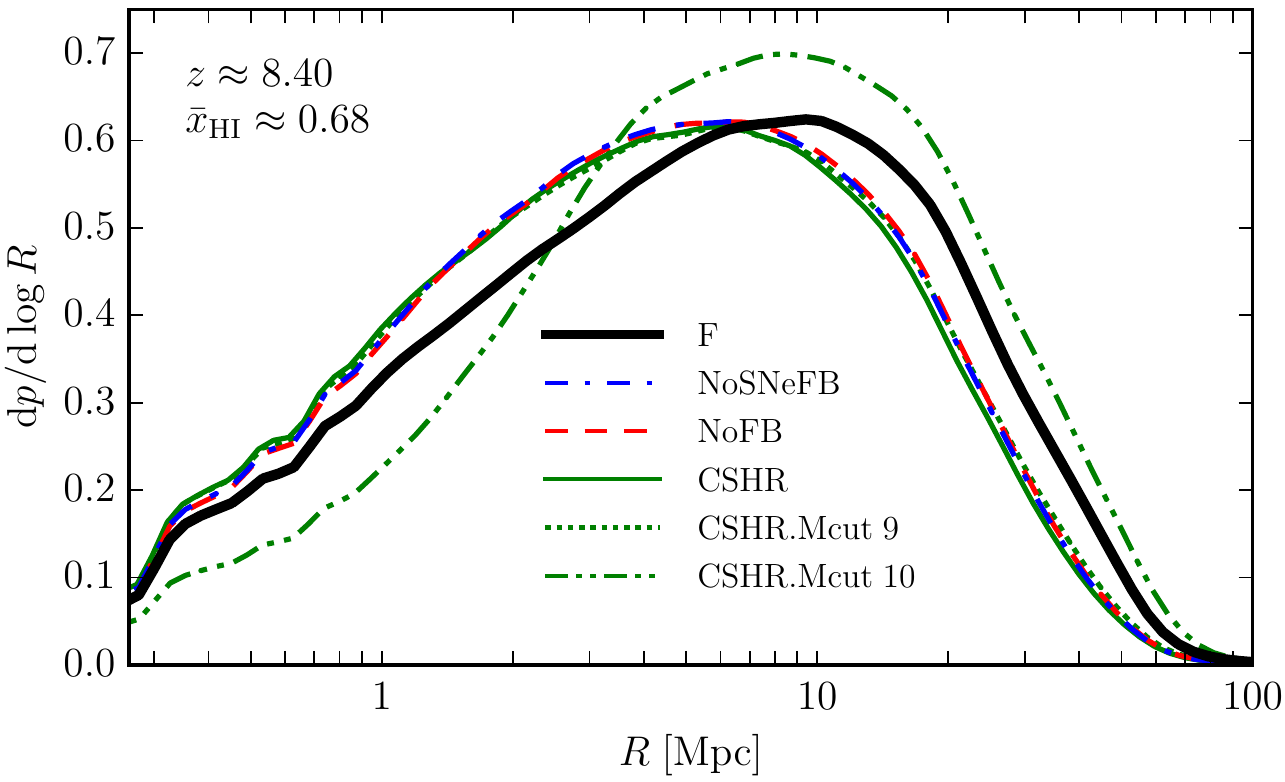}
\end{center}
\caption{Variation in the bubble size distribution with changes in the underlying astrophysics of the source model. Reproduced from \cite{Geil:2016}. Copyright of OUP Copyright 2019.}
\label{fig:BSD}
\end{figure}

In Figure~\ref{fig:BSD}, we highlight the observed variation in the bubble size distribution at essentially a fixed redshift/neutral fraction from \cite{Geil:2016}. The bubble size distributions here show the characteristic log-normal distribution, with the width of the peak and the relative extents of the asymmetric tails providing sufficient constraining information to distinguish between the various astrophysical models. While several curves appear to produce very similar bubble size distributions, folding in multiple epoch data should be enough to discriminate amongst various astrophysical parameters.

\subsection{Individual images}

Tomographic images of the 21-cm signal provide a direct tangible link to the process of reionisation, revealing the exact locations of ionised regions and potentially even directly observing the sources responsible with targeted follow up observations. For example, individual ionised regions can distinguish between ionisation driven by galaxies and quasars (e.g. \cite{Datta:2012,Majumdar:2012}), and also between other astrophysical sources such as galaxies containing either population II or III stars, mini-quasars, high-mass X-ray binaries or mini-haloes \cite{Ghara:2016,Ghara:2017}. This arises either directly from the size or shape of the ionised region (i.e. larger, more spherical regions in the case of AGN) or from the properties of the 21-cm signal in the immediate vicinity of the ionised region (i.e sharp or gradual changes in the 21-cm signal indicative of the spectrum of emitted ionising or X-ray radiation).

However, the signal-to-noise on a 21-cm image is considerably reduced as we cannot perform an averaging to boost the signal and further we observe the differential brightness temperature which is not necessarily a zero mean quantity, making the definition of an ionised (zero brightness temperature) region complicated. In order to counteract this, matched filters have been one proposed \cite{Datta:2007,Datta:2008,Datta:2012,Majumdar:2012,Malloy:2013,Datta:2016}, which act to minimise the contributions from the noise and foregrounds while maximising the signal by choosing a filter shape consistent with the expected feature of the signal (i.e. spherical ionised region). The 21-cm image is convolved with filters which vary in size and/or shape until the signal-to-noise of the product peaks. A peak in the signal-to-noise corresponds to a feature in the 21-cm image of the same shape as the filter. Matched filters have been explored both in the context of blind and targeted searches of ionised regions.

Alternatively, one can also extract information directly from a 21-cm image using machine learning techniques. Rather than searching for a specific feature (i.e. ionised region), a neural network can be constructed to perform a feature classification to identify regions of interest (see Section~\ref{sec:NN} for more details). Since the 21-cm data is in the form of a 2D (or 3D) image, the preferred network of choice is a convolutional neural network (CNN). The network is constructed using a training set of either 2 or 3D images (i.e. simulated images varying the astrophysical source properties), which undergo a series of down-samplings, convolutions and linear transformations which determine the weights for the various network layers that are used to identify specific features. The network architecture is both user and application specific, and will output user defined properties or parameters. Once the network is constructed, passing an image of the 21-cm signal to the network outputs the desired properties.

In recent years, the usage of CNNs have gained considerable traction. For example, \cite{Hassan:2019a} developed a CNN to distinguish between either AGN or galaxy driven reionisation, \cite{LaPlante:2018} extracted the global history of reionisation from their CNN, and both \cite{Hassan:2019b} and \cite{Gillet:2019} developed CNNs to extract astrophysical or cosmological parameters directly from the input 21-cm image.

\subsection{Stacked images}

Owing to the expected low signal-to-noise measurement for a 21-cm image, and that the individual ionised bubbles may be too small to be directly observed (compared to the resolution of the radio interferometer), \cite{Geil:2017} explored stacking redshifted 21-cm images centred on the known positions of high-redshift galaxies. Such an approach requires a precise determination of the galaxies redshifts on which the stack is centred otherwise the signal will be smeared out.

The resultant stack-averaging of the 21-cm signal produces a notably higher signal-to-noise detection for the mean ionisation profile by averaging out the statistical fluctuations within the IGM. If the IGM is in emission (i.e. heating has occurred), the stack averaged profile is observed in absorption. In contrast, if the IGM is in absorption (i.e. little to no heating) then the stack averaged profile is in emission. This stack-averaged profile then provides a rough estimate of the typical bubble size surrounding galaxies of known absolute UV magnitude which is important for determining if reionisation is driven by many small galaxies or larger, more biased galaxies. However, there remains a degeneracy between the bubble size and the ionisation state of the IGM. A stack of small ionised bubbles can be mimicked by a stack of larger ionised bubbles in a more ionised IGM (owing to the dependence of the mean 21-cm signal on the ionisation state of the IGM).

\subsection{Multi-field approaches}

Thus far we have discussed statistics purely focussed on the 21-cm signal. However, information can also be gleaned from combining the 21-cm signal with other independent tracers of the cosmological information. This can either be performed using a cross-correlation approach, where the 21-cm signal is cross-correlated with an alternative tracer of the galaxy or matter distribution. The advantage of this approach is that the foregrounds between these two fields should be completely uncorrelated, meaning they do not impact the underlying astrophysics of interest. Alternatively, a multi-tracer approach has been proposed, whereby the ratio of two measured fields (one being the 21-cm signal) are taken, which result in the underlying matter perturbations cancelling out leaving behind the interesting astrophysical information.

The leading example of the former approach is the cross-correlation between the 21-cm signal and Ly$\alpha$ emitting galaxies (LAEs; \cite{Wyithe:2007,Wiersma:2013,Sobacchi:2016,Vrbanec:2016,Heneka:2017,Hutter:2017,Hutter:2018a,Kubota:2018}). Here, the idea is that LAEs reside within the ionised regions, where the 21-cm signal is essentially zero (i.e. very little neutral hydrogen). Outside of these regions, the resonant scattering of the Ly$\alpha$ photons by the neutral hydrogen in the IGM strongly attenuates the Ly$\alpha$ line making these LAEs more difficult to detect, however the IGM is visible through the 21-cm signal. Thus, on radii smaller than the typical sizes of ionised regions the signal is anti-correlated. The anti-correlation then decreases to zero, or being slightly positive on much larger radii. The amplitude of this cross-correlation signal, and the rate at which the signal transitions from anti-correlation to zero can be used to determine the neutral fraction of the IGM as well as distinguishing different reionisation morphologies. Alternatives to LAEs have additionally been explored in the literature (\cite{Furlanetto:2007,Lidz:2009,Park:2014,Beardsley:2015}).

In the multi-tracer approach, two or more tracers of the same underlying field (i.e. the large-scale matter density) are used to extract astrophysical information. In taking the ratio of these fields, the matter density field cancels, leaving the astrophysics and cosmological terms. When using the 21-cm signal in combination with a field tracing the high-redshift galaxies, \cite{Fialkov:2019a} found that the anisotropy in the ratio can recover the sky-averaged 21-cm signal, distinguishing various models of the spectral energy distribution of the X-ray sources or the galaxy bias of the high-redshift galaxies. Importantly, in the absence an overlapping high-redshift galaxy survey, any alternative probe of the high-redshift universe can be used including for example planned CO or [CII] line intensity mapping of high-redshift galaxies (e.g. \cite{Kovetz:2017,Moradinezhad:2019a,Moradinezhad:2019b}).

\section{Modelling the 21-cm signal} \label{sec:models}

The 21-cm signal contains a wealth of cosmological and astrophysical information, too complex to be able to interpret without numerical methods. Our ability to learn about the underlying physical processes driving reionisation and the cosmic dawn hinges on being able to perform as physically accurate simulations as possible. However, such simulations require an enormous dynamic range, simultaneously resolving the small-scales (sub-kpc) in order to model the individual sources while also exploring the large-scale ($\sim100$'s of Mpc) radiative transfer effects of the high energy (e.g. X-ray) astrophysical processes responsible for heating and ionising the intergalactic medium. Further, in order to be able to produce an accurate representation of the observed 21-cm signal requires performing multiple simulations to explore the allowed parameter space.

In this section, we explore the various approaches taken within the literature to be able to simulate the 21-cm signal with the ultimate goal of learning as much about the underlying physics as possible. These will include describing the various existing approaches to simulate the 21-cm signal, while others will describe novel methods to inform where in parameter space to concentrate our efforts or methods to bypass performing the simulations all together.

\subsection{Numerical simulations}

Fully numerical simulations are designed to be the most physically accurate approach to investigate the underlying astrophysical processes. These generally consist of simulating the matter (baryons and dark-matter) either through N-body or hydrodynamical methods, and can additionally couple these with radiative transfer (either on-the-fly or post-processing) in order to simulate the radiation transport of the photons responsible for ionising/heating the IGM. The sheer complexity of the dynamic range required to accurately simulate the reionisation process often limits the physical volume of the simulation. However, through advances in computer design and processing power along with the ongoing development of more sophisticated computational algorithms we are continually able to push the boundaries with these types of simulations.

The most physically accurate approach is to perform full radiation hydrodynamical simulations capable of modelling the ionising sources and their interplay with the IGM (e.g. \cite{Ciardi:2001,Gnedin:2006,Finlator:2011,Gnedin:2014,Wise:2014,So:2014,OShea:2015,Norman:2015,Ocvirk:2016,Pawlik:2017,Ocvirk:2018,Rosdahl:2018,Wu:2019}). However, depending on the mass and spatial resolution of the small-scales these are very restrictive in their physical volume ($<100$~Mpc). A computationally cheaper approach is to couple a dark matter only or hydrodynamical simulation with coarser radiative transfer performed in post-processing (e.g. \cite{Iliev:2006,McQuinn:2007,Trac:2007,Ciardi:2012,Iliev:2014,Dixon:2016}). Such an approach enables notably larger simulation volumes to be explored ($<500$ Mpc) better suited for exploring the large-scale astrophysical processes, however, they typically require sub-grid modelling of the astrophysics.

It is through these classes of simulations where we will gain the largest insights into the astrophysical processes driving reionisation and the cosmic dawn. However, the computational cost of running these simulations is too prohibitive to perform a proper parameter exploration of the astrophysical processes. Thus, fully numerical simulations will need to be informed about interesting regions of astrophysical parameter space by analytic or semi-numerical simulations (Section~\ref{sec:efficientsims}).

\subsection{Semi-numerical and analytic models of the 21cm signal} \label{sec:efficientsims}

Rather than attempting to self-consistently model all the astrophysical processes, instead one can judiciously make a number of simplifying approximations in order to drastically increase the computational efficiency of the simulations. This can enable (i) huge cosmological volumes (several Gpc) and (ii) large numbers of simulations to be performed for rapid exploration of the astrophysical parameter space. It's with the approaches discussed below that a lot of progress can be made through being able to perform probabilistic searchers in the full astrophysical parameter space.

Semi-numerical simulations bypass radiative transfer all together, replacing it with an approximate scheme from which the ionisation field can be determined. One of the main approaches to do this is through the excursion-set approach (e.g. \cite{Furlanetto:2004}), which spatially distributes the ionising radiation by comparing the number of ionisations against recombinations in decreasing sized spherical shells (e.g. \cite{Mesinger:2007,Zahn:2007,Geil:2008,Alvarez:2009,Santos:2010,Mesinger:2011,Visbal:2012,Kim:2013,Fialkov:2014a,Majumdar:2014,Choudhury:2015,Hassan:2016,Kulkarni:2016,Mutch:2016,Hutter:2018b,Park:2019}). The determination of the number of ionising photons within each grid cell can either be determined from the underlying density field using excursion-set analytic halo mass functions or from identifying the discrete sources directly. Alternatively, one can calibrate a relation between the density field obtained from numerical simulations with properties of reionisation. For example, \cite{Battaglia:2013} use the relation between the redshift of reionisation and the bias of the underlying density field, while \cite{Kim:2016} use a relation between the ionisation fraction and the density.

Instead of bypassing the radiative transfer altogether, one can instead replace the three dimensional radiative transfer with a simple one dimensional radiative transfer and assume spherical symmetry for the distribution of the ionisation fronts (\cite{Thomas:2009,Ghara:2015}) to boost the computational efficiency of the simulations. 

Finally, if we are not interested in the three dimensional structure of reionisation we can construct simplified semi-analytic models which can describe the global history of reionisation and the cosmic dawn. Realistic reionisation histories can be obtained by solving the reionisation equation,
\begin{eqnarray} \label{eq:reion}
\frac{dQ}{dt} = \frac{n_{\rm ion}}{dt} - \frac{Q}{\bar{t}_{\rm rec}}
\end{eqnarray}
where $Q$ is the volume average filling factor of the Universe, $n_{\rm ion}$ is the number of ionising photons produced per baryon and $\bar{t}_{\rm rec}$ is the average recombination time-scale for neutral hydrogen. Using the excursion-set approach applied in one dimension (e.g. \cite{Furlanetto:2004}), we can determine the fraction of collapsed mass above some threshold level (barrier) given some mass threshold (e.g. halo mass). This analytic approach to solve Equation~\ref{eq:reion} has been extensively used in the literature as it gives a rapid and simple estimate of the number of ionising photons required to reionise the Universe \cite{Choudhury:2005,Choudhury:2006,Haardt:2012,Kuhlen:2012,Bouwens:2015,Mitra:2015,Robertson:2015,Khaire:2016,Madau:2017,Mitra:2018,Finkelstein:2019,Mason:2019,Naidu:2019}.

Extending from the excursion-set approach applied in one dimension (e.g. \cite{Furlanetto:2004}), other works have sought semi-analytic approaches to construct statistics describing the reionisation epoch. For example, \cite{Paranjape:2014} developed a model which provides expressions for the bubble-size distribution of the ionised regions, while \cite{McQuinn:2005,McQuinn:2006} explored analytic expressions to describe the 21-cm power spectrum (equivalently, \cite{Barkana:2007} explored the two dimensional correlation function). An alternative semi-analytic approach to describe the global 21-cm signal was developed by \cite{Mirocha:2014,Mirocha:2017,Mirocha:2018}.

\subsection{Intelligent sampling of the parameter space} \label{sec:intel_samp}

Understanding the astrophysics of reionisation and the cosmic dawn will require an exploration of astrophysical parameter space in order to be able to reveal the physical insights describing the observed 21-cm signal. Increasing the complexity, i.e. increasing the number of astrophysical processes or parameters that are simulated can make even these relatively computationally in-expensive semi-numerical simulations inefficient for parameter exploration. However, rather than exploring the entire astrophysical parameter space, we can instead make intelligent choices about which combinations of parameters we choose to sample within our simulations to minimise the computational costs. Such approaches can be useful for obtaining astrophysical parameter constraints directly (when combined with a metric such as a distance relation or likelihood which describes how well the model matches an observation), or for optimal designs for constructing training sets for machine learning approaches (see e.g. Section~\ref{sec:NN}). 

The most na\"{\i}ve approach is to construct a fixed grid of simulations, sampling evenly along each dimension of the astrophysical parameter space. However, as the number of dimensions increases, even this approach can become computationally intractable. An alternative approach is to consider sampling the parameter grid using a Latin-Hypercube approach (\cite{McKay:1979}). Here, the idea is to place points in the parameter grid to ensure no astrophysical parameter is sampled twice (see e.g. \cite{Kern:2017,Schmit:2018}). This approach minimises the overlap amongst the astrophysical parameters in the parameter set. Depending on our purpose, we can improve further on the Latin-Hypercube approach. If we have a reasonable idea with regard to the region of parameter space we expect the signal to occur, we can apply a spherical prior on the parameter space (e.g. \cite{Schneider:2011}). This sphericity drastically reduces the amount of volume in the hyper-surface that needs to be filled with samples (i.e. we ignore the edges of the parameter space). As we increase the dimensionality, the gains in reduction in volume become considerable (see e.g. the discussion in \cite{Kern:2017}).

Alternatively, rather than directly sampling the astrophysical parameter space by drawing from the astrophysical parameters, one can instead adopt a Jeffreys prior (\cite{Jeffreys:1946}). Such an approach searches for regions of the parameter space where the observable (e.g. statistic of the 21-cm signal) varies maximally and increases the sampling within such a region, producing coarser sampling elsewhere. \cite{Eames:2018} explored usage of this Jeffrey's prior in sampling the astrophysical parameter space for reionisation simulations. In addition to the Latin-Hypercube approach described above, they also explored sampling the parameter space using average-eigenvector sampling and adaptive grid-free sampling. The latter two use a hyper-surface distance based metric to inform the placement of points of interest in the parameter space. Such approaches can drastically improve the performance of neural network based approaches by ensuring optimal designs for the training sets (see Section~\ref{sec:NN}).

\subsection{Emulators} \label{sec:emul}

If we are only interested in a statistical description of the 21-cm signal (e.g. 21-cm power spectrum), we can bypass performing the entire numerical or semi-numerical simulations in favour of constructing an emulator. Emulators are a machine learning technique which aims to replicate the desired output of a model using either a series of functional curves (for example polynomials) or a neural network. Once constructed, the emulator provides the desired output statistic describing the signal almost instantaneously given a set of input astrophysical parameters, which can drastically improve parameter exploration. Emulators have been used within astrophysics for a while (see e.g. \cite{Heitmann:2009,Agarwal:2012,Heitmann:2014,Heitmann:2016}), however, only recently have they been explored in the context of reionisation. The construction of an emulator benefits from the intelligent sampling of the astrophysical parameter space (e.g. Section~\ref{sec:intel_samp}) to minimise the size of the required training set.

\cite{Kern:2017} constructed a Gaussian Process (GP) based emulator of the 21-cm power spectrum for the semi-numerical simulation code 21cmFAST \cite{Mesinger:2007,Mesinger:2011}. This emulator takes as input 11 parameters, 5 cosmological and 6 astrophysical, and outputs the 21-cm power spectrum at any redshift during the  reionisation and cosmic dawn epochs. In order to accelerate the training and construction of the emulator, rather than using the raw 21-cm power spectrum outputs ($\Delta^{2}_{21}(k,z)$) a data compression step can be performed to minimise the number of features that the emulator needs to learn. In this work, a principal component analysis (PCA) approach was adopted, which minimises the number of independent pieces of information required to describe the 21-cm power spectrum (i.e. replace the full correlated $k$-bin range, with the sum of a few PCA components). The emulator is then constructed using GP regression to minimise a GP generator function which is completely defined by its mean and covariance within the astrophysical parameter space. For further details, refer to \cite{Kern:2017}.

Alternatively, \cite{Schmit:2018} construct an emulator of the 21-cm power spectrum from 21cmFAST using an artificial neural network (see Section~\ref{sec:NN} for further details and applications). This neural network takes as input the astrophysical parameters describing the model and directly returns an estimate of the 21-cm power spectrum. Evaluating the network to obtain a new 21-cm power spectrum is effectively instantaneous. 

Further, \cite{Jennings:2019} explored several different possible techniques to construct an emulator of semi-numerical simulations. In addition to two simplistic interpolation techniques (i.e. interpolate the result between points in the training set of data), they also explored neural networks, GPs and a support vector machine (SVM). They find that a neural network approach performs best (e.g. \cite{Schmit:2018}) however note that the more sophisticated GP and SVM approaches could be optimised to outperform a neural network emulator.

Instead of simply emulating a function describing the 21-cm signal statistics, recently \cite{Chardin:2019} developed an emulator for the radiative transfer process within reionisation simulations. This approach uses deep learning (another machine learning technique) to output three-dimensional maps of the reionisation time in each cell given an input two dimensional map of the number density of stars and gas. Specifically it uses a trained auto encoder convolutional neural network, which uses layers of two-dimensional convolution kernels to describe the system that is being emulated.

\subsection{Characterising our ignorance}

The trade-off for increased computational efficiency with the semi-analytic and semi-numerical approaches described in Section~\ref{sec:efficientsims} is the reduction in numerical accuracy. When it comes to extracting information about the astrophysics of reionisation and the cosmic dawn from the 21-cm signal, we must therefore be fully aware of the shortcomings of our simulations in order to be able to interpret the results. Further, we must understand in what regimes we can confidently trust these approximate simulations.

For example, in \cite{Santos:2008} some of the analytic models described in Section~\ref{sec:efficientsims} were compared against a hybrid N-body and radiative transfer simulation of cosmic reionisation. Under certain regimes, these analytic models are shown to perform relative well at matching the statistics of the 21-cm power spectrum. Following on from this, \cite{Zahn:2011} explored the comparison between radiative transfer simulations and semi-numerical simulations. In terms of the 21-cm power spectrum, differences between these approaches were found to be of order of 10 per cent in the power spectrum amplitude. With these sorts of comparisons, we can gain confidence that parameter explorations using approximate techniques can reveal useful astrophysical insights. However, for these algorithm comparisons only single astrophysical models were considered. To truly characterise our ignorance a larger, more detailed suite of simulations would be required to fully ascertain how good an approximation they are.

This tens of per cent level uncertainty can be added as an additional modelling uncertainty in attempts to recover astrophysical parameters from the 21-cm signal. This effectively acts as an uncertainty floor, with parameter constraints only available where the impact of the astrophysical parameters on the 21-cm signal is larger than this modelling uncertainty. \cite{Greig:2015,Greig:2019} adopt a 25 per cent modelling uncertainty error to the 21-cm power spectrum finding no biases in the recovery of the astrophysical parameters.

Semi-numerical simulations based on the excursion-set formalism are explicitly photon non-conserving. That is, not all ionising photons are exhausted (i.e. they are lost to the ether) within the simulation. The basis of this, is that the analytic solutions from excursion-set theory are photon conserving in one dimension, however, the three dimensional application in semi-numerical simulations is not. When bubble overlap occurs, ionising photons are not redistributed from the overlap region, they are just unused. This photon non-conservation can result in notably biases in the amplitude of the 21-cm power spectrum \cite{Choudhury:2018} when comparing simulation outputs. However, this photon non-conservation can be trivially accounted for by rescaling the production rates of ionising photons to match the expected global reionisation histories (e.g \cite{Zahn:2011,Majumdar:2014}. Alternatively, more robust corrections can be considered (e.g. \cite{Paranjape:2016,Choudhury:2018,Molaro:2019}).

Provided we are aware of the issues and account for the biases or limitations of the approximate schemes, in most cases we should be able to confidently use these approaches for detailed astrophysical parameter exploration. Of course in practise it is difficult to verify this for each possible astrophysical model or summary statistic.

\section{Inference methods for the 21-cm signal}

In the previous chapters we have discussed the astrophysics and cosmology encoded within the 21-cm signal. In Section~\ref{sec:methods} we discussed numerous ways to characterise the 21-cm signal to tease out the interesting astrophysics, while in Section~\ref{sec:models} we discussed the various approaches to model the 21-cm signal. The final piece to unlocking the astrophysical information from a 21-cm observation is through performing a robust probabilistic exploration of our simulated astrophysical parameter space. This requires comparing the observed 21-cm signal (or a statistic characterising it) against the synthetic output from our simulations, taking into account all forms of possible uncertainties (both observational and theoretical). 

Ultimately, we are interested in obtaining the probability distribution function (PDF) of the entire astrophysical parameter space from our simulated model (or the posterior distribution, $P(\mathbf{\theta}|\mathbf{d})$; the probability of the model astrophysical parameter set, $\mathbf{\theta}$, given the observational data, $\mathbf{d}$). This is what is referred to as the posterior distribution which is obtained from Bayesian statistics through Bayes' theorem,
\begin{eqnarray} \label{eq:Bayes}
P(\mathbf{\theta}|\mathbf{d}) = \frac{P(\mathbf{\mathbf{d}|\theta})P(\mathbf{\theta})}{P(\mathbf{d})},
\end{eqnarray}
where $P(\mathbf{\mathbf{d}|\theta})$ is the likelihood which describes how likely the astrophysical model described by the parameter set $\theta$ describes the data, $P(\mathbf{\theta})$ contains all the prior information we have about the specific astrophysical parameters within our model and $P(\mathbf{d})$ is the evidence which measures how likely the data is given the model. Throughout this section, we will discuss the various approaches considered in the literature for obtaining the posterior PDF. 

\subsection{Fisher Matrices}

One of the simplest and easiest to implement approaches to obtain astrophysical constraints is from the Fisher information matrix (\cite{Fisher:1935}, see e.g. \cite{Tegmark:1997,Coe:2009} for examples how to implement it). This provides a method to quantify the amount of information that an observation contains about any of the unknown parameters in the model parameter set, $\mathbf{\theta}$. The Fisher information matrix, $\mathbf{F}$, is calculated via,
\begin{eqnarray}
\mathbf{F}_{ij} = \left\langle \frac{\partial^{2}{\rm ln}\mathcal{L}}{\partial\theta_{i}\partial\theta_{j}} \right\rangle = \sum_{\mathbf{x}} \frac{1}{\epsilon^{2}(\mathbf{x})} \frac{\partial f(\mathbf{x})}{\partial\theta_{i}} \frac{\partial f(\mathbf{x})}{\partial\theta_{j}} ,
\end{eqnarray}
where $\mathcal{L}$ is the likelihood function (probability distribution of the observed data given the astrophysical parameter set) and $\epsilon$ characterises the error on the measurement of the function, $f(\mathbf{x})$, where $\mathbf{x}$ is the data vector describing the function (i.e. for the 21-cm power spectrum this would be $(k,z)$ the Fourier wavenumber, $k$, and the redshift, $z$). Here, $\mathbf{\theta}$ is the astrophysical parameter set, and we sum the contribution of the partial derivatives of the measured function with each parameter. Parameters which result in large variations in the partial derivatives contain considerable weight and thus highlight which model parameters are sensitive to the function describing the observational data.

Evaluating the Fisher matrix firstly requires the determination of the maximum likelihood model. We can either assume a fiducial parameter set maximises the model, or we can find the model parameter set which is maximal given the observational uncertainties. In the latter case, this can be somewhat computationally expensive, as it requires determining the maximum of our likelihood function.

Once the Fisher information matrix has been calculated, the resultant errors on the model parameters, $\mathbf{\theta}$, given the observation can be obtained by inverting $\mathbf{F}_{ij}$. That is,
\begin{eqnarray}
\mathbf{C}_{ij} = \frac{1}{\mathbf{F}_{ij}},
\end{eqnarray}
where $\mathbf{C}$ is the covariance matrix, with the diagonal entries, $\mathbf{C}_{ii}$, containing the errors on the model parameters (i.e. $\mathbf{C}_{ii} = \sigma^{2}_{ii}$, where $\sigma$ is the standard deviation), and the off-diagonal entries describing the two-dimensional joint probabilities which highlights the degeneracies between those two specific model parameters (i.e. how much similar information each parameter holds).

Fundamentally, the Fisher matrix approach assumes that the observation has been performed optimally, where the uncertainty, $\epsilon$, contains a full description of all sources of error. Further, the inversion of the Fisher matrix to obtain parameter uncertainties assumes that the model parameter set is fully described by a Gaussian likelihood (which is rarely the case in reality). Despite these short-comings, the Fisher matrix provides an excellent and computationally efficient means to provide astrophysical parameter constraints given an observation of the 21-cm signal.

Forecasting of astrophysical or cosmological parameters from during reionisation and the cosmic dawn using Fisher matrices has been extensively used in the literature. For example, with the 21-cm power spectrum \cite{Pober:2014} explored the forecasts for parameters responsible for reionisation, \cite{Liu:2016b} explored similar parameters but coupled with cosmological parameters, while \cite{Ewall-Wice:2016} considered the astrophysical parameters responsible for X-ray heating. Alternatively, \cite{Kubota:2016} explored astrophysics from the variance and skewness of the 1D PDF of 21-cm fluctuations while \cite{Shimabukuro:2017} instead investigated the 21-cm bispectrum and \cite{Pritchard:2010a} explored the global 21-cm signal. Pure cosmology or joint cosmological and astrophysics were additionally investigated using analytic expressions of the reionisation epoch by \cite{McQuinn:2006,Mao:2008,Barger:2009,Visbal:2009,Liu:2016b}.

\subsection{Fixed grid sampling}

The simplest approach to recover the true PDF of our astrophysical parameters (i.e. not under the Gaussian approximation applied in the case of the Fisher matrix) is to construct a grid of astrophysical models which are sampled along the dimensions of the allowed astrophysical parameters. This can either be in a fixed, evenly sampled grid along each dimension or a more informed grid sampling as discussed in Section~\ref{sec:intel_samp} which reduces the number of models required. Once the grid has been constructed, at each grid point we then compare the observed 21-cm signal against the simulated output given that set of astrophysical parameters to assign it a probability (e.g. the likelihood of it being the correct description of the observed data). With this grid of probabilities, we can then interpolate it to generate a full (continuous) description of the underlying PDF (e.g. $P(\mathbf{\theta}|\mathbf{d})$) for this specific astrophysical setup. With this PDF, we can then obtain constraints on any specific astrophysical parameter within our model by marginalising (integrating) over the uncertainties in all other parameters;
\begin{eqnarray}
P(\theta_{1}|\mathbf{d}) = N^{-1} \int P(\mathbf{\theta}|\mathbf{d}) d\theta_{2},d\theta_{3},...,d\theta_{n},
\end{eqnarray}
where $\mathbf{\theta} = (\theta_{1}, \theta_{2}, ..., \theta_{n})$ is the astrophysical parameter set and $N$ is the normalisation constant which ensures $\int P(\mathbf{\theta}|\mathbf{d}) d\mathbf{\theta} = 1$.

Grid based sampling of astrophysical models has been used throughout the literature both for parameter forecasting, as well as inference from observed upper limits on the 21-cm signal. For example, limits on astrophysical parameters during reionisation and the cosmic dawn using the 21-cm global signal have been explored with the Experiment to Detect the Global EoR Signature (EDGES; \cite{Monsalve:2017,Monsalve:2018,Monsalve:2019}) as well as the Long Wavelength Array (LWA; \cite{Fialkov:2019b}). Grids of semi-numerical simulations of reionisation have also been used to interpret existing constraints on reionisation (e.g. such as the optical depth, $\tau_{e}$ \cite{Planck:2018} or limits on the IGM neutral fraction \cite{McGreer:2015}) in the context of PDFs of astrophysical model parameters (e.g. \cite{Mesinger:2012,Mesinger:2013,Greig:2017a}). The equivalent has also been considered for analytic methods \cite{Choudhury:2005,Barkana:2009,Zahn:2012,Mirocha:2018}.

While the fixed grid approach recovers the true underlying PDF and thus is more accurate than the Fisher matrix, it is considerable more computationally expensive due to the increase in number of simulations required. Further, this assumes that the likelihood space is well behaved and varies smoothly. If the likelihood varies sharply, then finer resolution sampling would be required around those regions of parameter space. It is tractable for a low number of astrophysical parameters, but once this goes beyond just a few free parameters it can become infeasible. In the next few sections we discuss techniques to circumvent this.

\subsection{Bayesian MCMC} \label{sec:MCMC}

Once the dimensionality of our astrophysical parameter space becomes too large to directly sample, we must shift to more approximate methods to recover the true PDF of our astrophysical model. In statistics, this is achieved through Markov-Chain Monte-Carlo (MCMC) methods, where we can obtain an estimate of our posterior distribution, $P(\mathbf{\theta}|\mathbf{d})$ through random sampling. To demonstrate the basic idea, we outline one of the simplest MCMC approaches, the Metropolis-Hastings algorithm (\cite{Metropolis:1953,Hastings:1970}). We start with a set of initial positions within our astrophysical parameter space, compute the product of the prior and the likelihood (e.g. the numerator of Equation~\ref{eq:Bayes}) and then take a random jump to a new position in the probability space. In this new position, we compute the product of the prior and likelihood corresponding to the new position, and compare against the previous position. If the new quantity is higher, we keep the new parameter set, if lower, we keep it some fraction of the time according to a probability check (e.g. generate a random number between zero and one and if its higher than the ratio (which is less than one), we keep it). Following this procedure through a large number of iteration, eventually the chain will converge to the peak of the posterior distribution as it must move to regions where the likelihood is higher (i.e. higher probability). To ensure robustness of the sampling (i.e. avoid local minima in our probability space), we perform many Markov-Chains. Once this is complete, simply constructing a histogram of all the sampled points returns an estimate of the posterior distribution (as the most frequent datapoints in the parameter space are those in regions of higher likelihood).

Returning to Bayes' theorem (Equation~\ref{eq:Bayes}), all we require for performing an MCMC is the likelihood ($P(\mathbf{\mathbf{d}|\theta})$) and the prior information ($P(\theta)$). Since at all points within the MCMC we take the ratio of the likelihood multiplied by the prior, we never require the evidence ($P(\mathbf{d})$). This is the advantage of the MCMC approach, as it is the evidence that is the most computationally expensive component of Bayes' theorem (to calculate the evidence we need to perform a multi-dimensional integral over our entire astrophysical parameter space). The last remaining component is the MCMC sampler itself, which is how to determine the new position within the MCMC chain. Within the field of statistics there are many difference types of MCMC samplers all with their own pros and cons, with plenty of literature to assist in deciding which approach is most suitable given the problem.

Over the past decade, the usage of MCMC techniques for the reionisation and the cosmic dawn have been gaining considerable attention. For analytic models that generate reionisation histories that can be coupled with CMB data and other observational constraints, MCMC approaches have been well established (e.g. \cite{Pritchard:2010b,Clesse:2012,Morandi:2012,Mitra:2015,Gorce:2018,Finkelstein:2019,Mason:2019,Naidu:2019}). Analytic models of the global 21-cm signal have also been explored with MCMC, both for interpreting observational limits and also for parameter forecasting for future 21-cm experiments (e.g. \cite{Pritchard:2010a,Harker:2012,Mirocha:2015,Bernardi:2016,Harker:2016}).

Only relatively recently have computational resources become efficient enough to be directly applied to semi-numerical simulations of the 21-cm signal (e.g. \cite{Greig:2015,Greig:2017b,Greig:2018,Greig:2019,Park:2019}). That is, to be able to perform a three dimensional simulation of the 21-cm signal at each set within the MCMC. Alternatively, one can also interpolate over a fixed grid of simulations within an MCMC framework (\cite{Hassan:2017}). Emulators can additionally be coupled to MCMC techniques, whereby the semi-numerical simulation is bypassed with an emulated function describing the 21-cm signal (e.g. \cite{Kern:2017,Schmit:2018}), drastically increasing the computational efficiency. An alternative hybrid approach is to instead train an emulator during the MCMC, to decide whether a new parameter position is close enough to previously sampled positions from which we can emulate the expected result or whether we need to perform the actual likelihood (simulation) call (e.g. \cite{vanderVelden:2019}).

\subsection{Model selection and nested sampling}

Let's return to Bayes' theorem (Equation~\ref{eq:Bayes}), and instead write it explicitly as a function of our chosen astrophysical model, $\mathbf{M}$,
\begin{eqnarray} \label{eq:BayesMS}
P(\mathbf{\theta}|\mathbf{M},\mathbf{d}) = \frac{P(\mathbf{\mathbf{d}|\mathbf{M},\theta})P(\mathbf{\theta}|\mathbf{M})}{P(\mathbf{d}|\mathbf{M})}.
\end{eqnarray}
All terms remain as in Equation~\ref{eq:Bayes}, however, let's focus explicitly on the Bayesian evidence, $P(\mathbf{d}|\mathbf{M})$ (which is often expressed as $\altmathcal{Z}$). This evidence quantifies how likely the observation data was, given our astrophysical model. In traditional MCMC techniques (see previous section), the evidence is ignored as it can be computationally expensive to evaluate and also it is a redundant calculation as we consistently take the ratio of the likelihood to estimate our positions in our astrophysical parameter space. However, if we instead evaluate the evidence term, $\altmathcal{Z}$, we can use this to perform model selection amongst a variety of potentially plausible astrophysical models (i.e. determine which model provides a better representation of the observational data). Herein lies the value of estimating the Bayesian evidence.

Model selection can be performed by taking the ratio of the Bayesian evidence for each model (known as the Bayes factor),
\begin{eqnarray} \label{eq:BayesMS}
\altmathcal{B}_{12} = \frac{P(\mathbf{d}|\mathbf{M}_{1})}{P(\mathbf{d}|\mathbf{M}_{2})}.
\end{eqnarray}
The Bayes factor informs us of the strength of the evidence for one model against another. In other words, if $\altmathcal{B}_{12}$ is greater than unity, then the evidence suggests that model 1 is better than model 2 at modelling our observational data. In \cite{Jeffreys:1961} the Jeffreys scale was introduced to provide a means to classify how strongly the evidence for one model is relative to another. Since then, this scale has been modified several times meaning there is no unique criterion. Here, we adopt the scaling provided by \cite{Lee:2014} which is broken down as follows: (i) if $\altmathcal{B}_{12} > 100$, there is extremely strong evidence for model 1 compared to model 2, (ii) if $30 < \altmathcal{B}_{12} < 100$ there is very strong evidence, (iii) if $10 < \altmathcal{B}_{12} < 30$ there is strong evidence, (iv) if $3 < \altmathcal{B}_{12} < 10$ there is moderate evidence, (v) if $1 < \altmathcal{B}_{12} < 3$ there is anecdotal evidence and (vi) $\altmathcal{B}_{12} = 1$ there is no evidence.

Model selection in the context of reionisation simulations has only relatively recently been explored in \cite{Binnie:2019}. Here, various semi-numerical simulations of reionisation were explored within the context of a mock 21-cm observation. The models differ in how reionisation proceeded (i.e. inside-out compared to outside-in) along with simpler prescriptions for simulating reionisation (i.e. excursion-set compared to a simpler pixel-by-pixel definition \cite{MiraldaEscude:2000}). Using model selection, certain models could be ruled out with mock observations from next generation radio interferometers.

In order to be able to perform model selection, we must be able to compute the Bayesian evidence. This can be achieved using the nested sampling algorithm (e.g. \cite{Skilling:2004}) which performs transformations of the astrophysical parameter space to collapse the multi-dimensional integral for the evidence into a series of more computationally feasible one-dimensional integrals. A convenient byproduct of the nested sampling algorithm is that in order to compute the evidence, one generates samples from the posterior distribution (e.g. $P(\mathbf{\theta}|\mathbf{M},\mathbf{d})$), thus it can perform the same task as a traditional MCMC algorithm. In fact, the particular approach to sampling the astrophysical parameter space within nested sampling can be notably more efficient in regard to the number of required model calls to estimate the posterior distribution. As such, nested sampling is often preferred over traditional MCMC algorithms.

\subsection{Neural Networks} \label{sec:NN}

We have already briefly touched upon neural networks (see Section~\ref{sec:emul}), however, that was in the context of constructing an emulator of the 21-cm power spectrum given input astrophysical parameters. Instead, in this section we flip the problem around and focus on the usage of neural networks to recover estimates of the underlying astrophysical parameters given some input observational dataset. That is, bypass MCMC techniques all together and infer astrophysics directly from a neural network. There have been several works in the literature exploring the validity of using neural networks to perform this task, and we will touch upon the similarities and differences of each of these different approaches.

\begin{figure}[]
\begin{center}
\includegraphics[trim = 0.2cm 1cm 0.2cm 0.2cm, scale = 0.45]{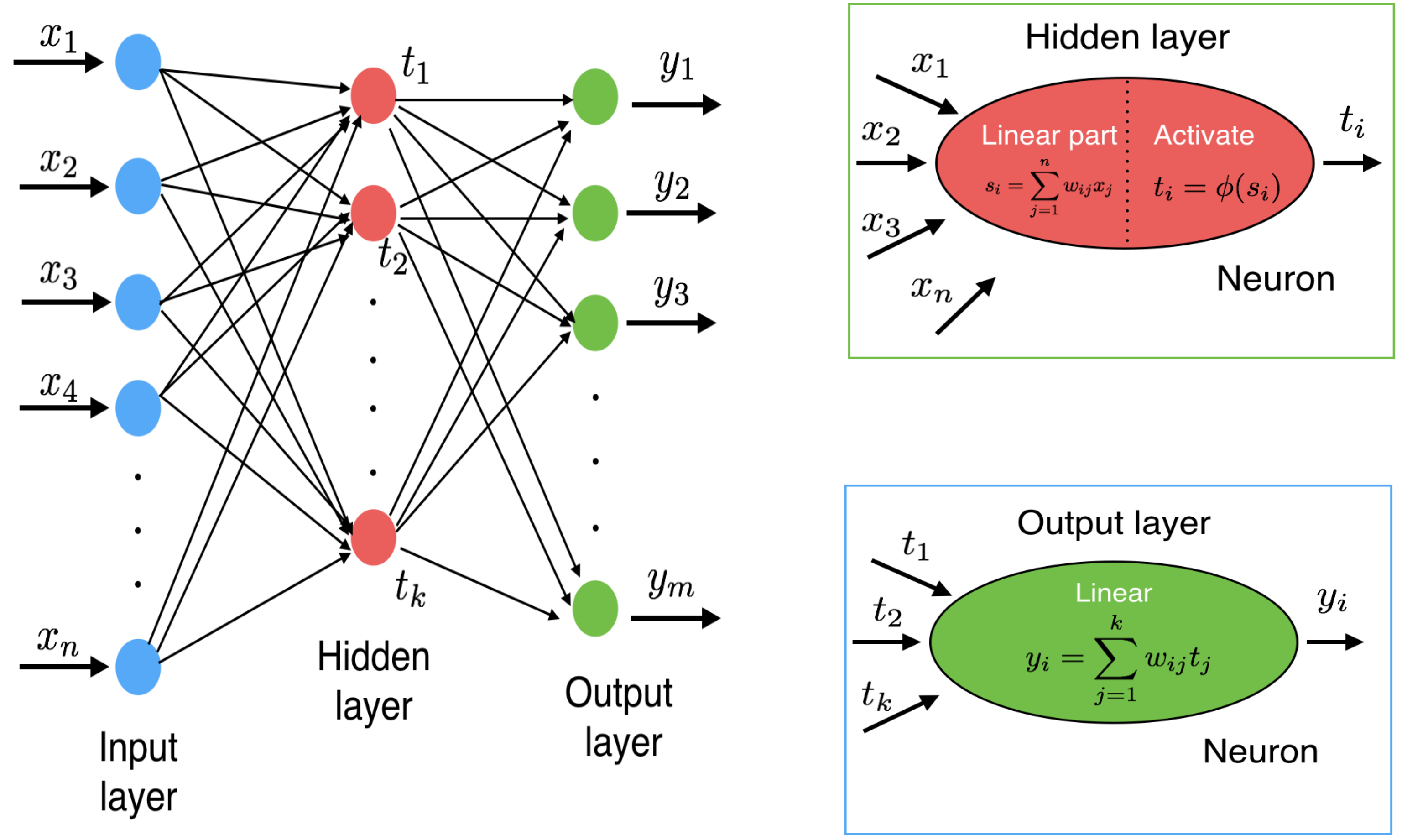}
\end{center}
\caption{An example architecture of an artificial neural network. Reproduced from \cite{Shimabukuro:2017b}. Copyright of OUP Copyright 2019.}
\label{fig:ANN}
\end{figure}

The fundamental idea of a neural network is to construct a computing system which mimics the behaviour of the brain, containing multiple layers of neurons (see Figure~\ref{fig:ANN} for an example). A neural network must contain an input layer (which processes the input data), any number of hidden layers and a final output layer which produces the desired user defined output (i.e. astrophysical parameters). Each neuron in a layer is connected to all other neurons in adjoining layers (it is not connected within its own layer) and the strength of the connection is driven by what is referred to as the activation function (which takes as input the sum of all results from all other neurons multiplied by the weight of each). We must then train the neural network (the weights returned by each neuron) given the inputs and outcomes of the training set we seek to learn. In order to be able to learn the weights, whats referred to as back propagation is often applied. Here, the aim is to estimate the weights for the network by minimising a cost function. This is an iterative procedure requiring many epochs, and we must be careful not to overfit our network. Thus, the number of epochs is not pre-determined but instead is typically taken to be the value when the cost-function first begins to plateau. We then validate the accuracy of the neural network by comparing the expected outputs from a new dataset (validation dataset, which must differ from the training set) against the returned output from the neural network. Once constructed and validated, the network then almost instantaneously returns the desired user defined outputs given the preferred input format.

\cite{Shimabukuro:2017b} explored the usage of artificial neural networks (ANN) in the context of astrophysical parameter recovery from the 21-cm power spectrum. The network was constructed to take as input a training set of 70 21-cm power spectra varying three astrophysical parameters, and return the expected value given an input 21-cm power spectrum. \cite{Doussot:2019} significantly improved upon this initial ANN approach, considering both a larger training set for the same astrophysical model (2400 models) and supervised learning techniques (techniques to improve the accuracy and optimisation of the constructed neural network, see \cite{Doussot:2019} for more details).

\begin{figure}[]
\begin{center}
\includegraphics[trim = 0.2cm 1cm 0.2cm 0.2cm, scale = 0.35]{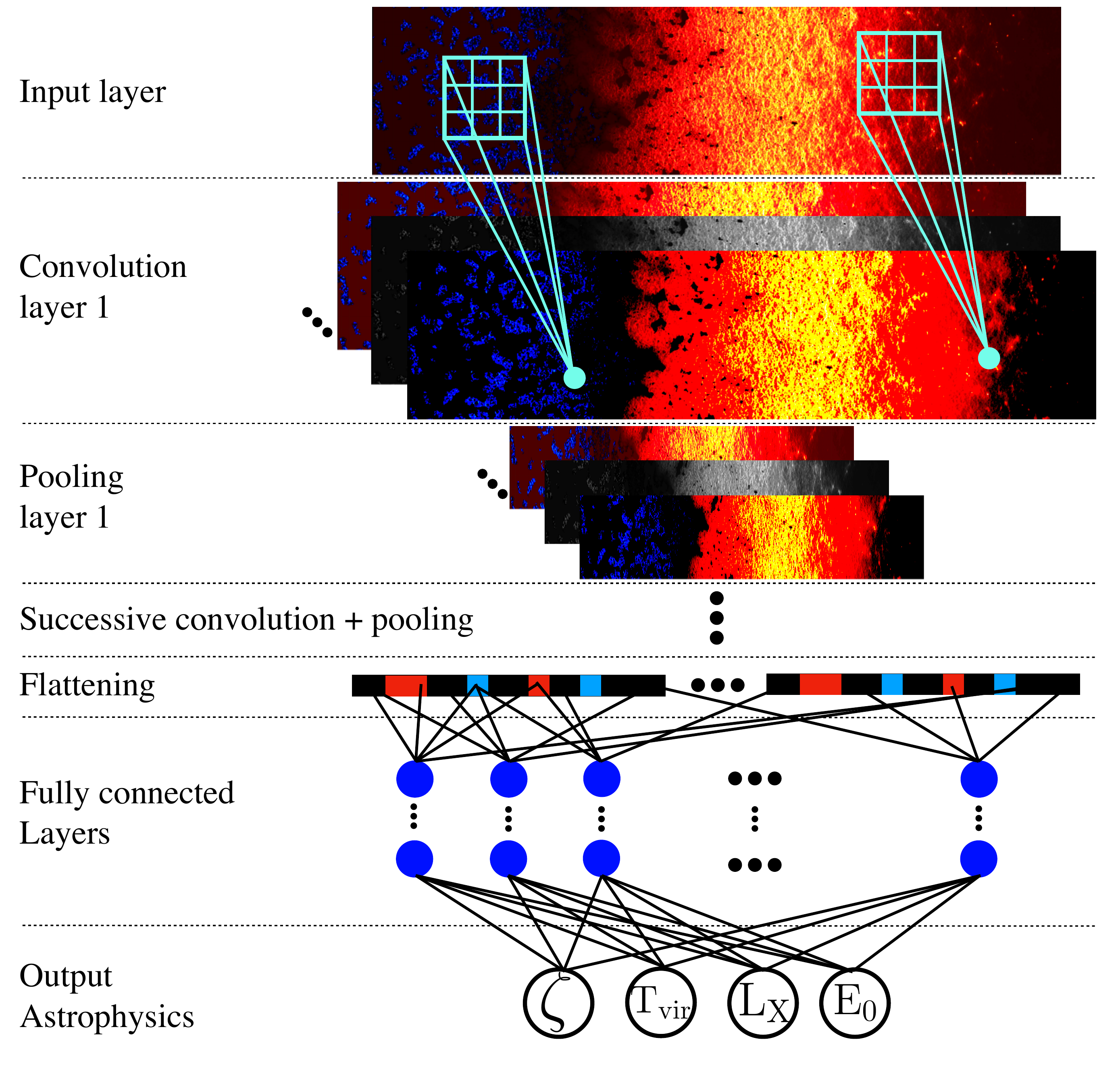}
\end{center}
\caption{An example architecture of a convolutional neural network. Showing the architecture from an input two-dimensional light-cone of the 21-cm signal down to the output astrophysical parameters. Reproduced from \cite{Gillet:2019}. Copyright of OUP Copyright 2019.}
\label{fig:CNN}
\end{figure}

Rather than only using a statistical descriptor of the 21-cm signal (i.e. power spectrum), we could instead use the expected full two or three dimensional 21-cm signal. To do this, we use a convolutional neural network (CNN), whose network architecture is designed to work with images. The main differences between an ANN and CNN is that the CNN requires feature extraction in order to break down the volume of data into a more manageable set. Feature extraction is performed by a series of convolutional and pooling layers on the input image, with each convolutional layer convolving the result with a number of filters to break down the image (see Figure~\ref{fig:CNN} for an example of a CNN) into a simpler set of values to be passed to the neurons of the network. Following feature extraction, the network of neurons is constructed in a similar fashion as in the case of an ANN.

CNNs have been used in a few different ways for reionisation and the cosmic dawn. For example, \cite{Gillet:2019} used two-dimensional light-cones of the simulated 21-cm signal in order to extract the underlying eight astrophysical parameters (from reionisation and X-ray heating) from a mock observation. Similarly, \cite{Hassan:2019b} jointly constrained three cosmological parameters along with three astrophysical parameters from reionisation using two-dimensional images of the 21-cm signal at several redshift snapshots.  Alternatively, \cite{LaPlante:2018} explored recovering the reionisation history from foreground dominated two-dimensional 21-cm images at several redshifts while \cite{Hassan:2019a} used a CNN to perform image classification to determine whether features in the 21-cm image could be distinguished as being driven by galaxies or active galactic nuclei.

Neural networks have been shown to perform extremely well at recovering the expected astrophysics from mock observations, and the evaluation of the neural network for parameter estimation is more computationally efficient than MCMC techniques\footnote{Although the construction of the training set can be as slow or slower than an MCMC, the training set may only need to be constructed once whereas an MCMC must be performed each time.}. However, the fundamental issue with neural networks is their relative inability to provide meaningful uncertainties on the recovered parameters. That is, to characterise how inherent uncertainties in the network construction propagate through into the recovered astrophysical constraints. Contrast this to the recovered posterior distributions from MCMC techniques. Nevertheless, neural networks are an extremely useful and valuable tool for inferring astrophysics about the reionisation and cosmic dawn.

\bibliographystyle{plain}
\bibliography{References}

\end{document}